\documentclass{article}
\usepackage{graphicx} 
\usepackage[english]{babel}    
\usepackage[utf8]{inputenc} 
\usepackage[T1]{fontenc}
\usepackage{amsmath, amssymb, amsthm}
\usepackage[libertine,bigdelims,vvarbb,scaled=1.05]{newtxmath} 
\usepackage[spacing=true,kerning=true,babel=true,tracking=true]{microtype}
\usepackage[hang, small, labelfont=bf, up]{caption}
\usepackage{booktabs} 
\usepackage{lastpage}
\usepackage[hidelinks]{hyperref}
\usepackage{color}
\usepackage{float}
\usepackage{mathtools}
\usepackage{tikz, subcaption}
\usepackage{multirow}
\usepackage{booktabs}
\usepackage{lipsum} 
\usepackage{fancyhdr}
\usepackage{geometry}
\usepackage{braket}
\usepackage{placeins}
\usepackage{changes}
\usetikzlibrary{shapes.geometric}

\newtheorem{theorem}{Theorem}[section]

\theoremstyle{definition}
\newtheorem{definition}[theorem]{Definition}

\theoremstyle{remark}

\usepackage{cleveref} 

\crefalias{theorem}{theorem}
\crefalias{definition}{definition}
\crefalias{corollary}{corollary}
\crefalias{lemma}{lemma}

\crefname{theorem}{Thm.}{Thms.}
\Crefname{theorem}{Theorem}{Theorems}

\crefname{definition}{Def.}{Defs.}
\Crefname{definition}{Definition}{Definitions}

\crefname{equation}{Eq.}{Eqs.}
\Crefname{equation}{Equation}{Equations}

\crefname{section}{Sec.}{Secs.}
\Crefname{section}{Section}{Sections}

\crefname{figure}{Fig.}{Figs.}
\Crefname{figure}{Figure}{Figures}

\crefname{table}{Tab.}{Tabs.}
\Crefname{table}{Table}{Tables}

\title{On Relationship Between Circuit Depth and Trainability of VQAs}

 \author{Jan Michálek$^{1}$, 
 Martin Friák$^{2}$, Petr Vašík$^{1}$\textsuperscript{*}\\
 \\
 $^{1}$ Brno University of Technology, Faculty of Mechanical Engineering, Czech Republic \\
 $^{2}$ Institute of Physics of Materials, Czech Academy of Sciences, v. v. i., Czech Republic \\
 \texttt{michalek.j@fme.vutbr.cz, friak@ipm.cz,} \\
 \texttt{ Petr.Vasik@vutbr.cz}
 }

\begin{document}

\date{}
\maketitle

\noindent
\textsuperscript{*}Corresponding author: Petr Va\v s\' ik\\

{\bf Abstract}\\[1mm]

The training efficiency of Variational Quantum Algorithms (VQAs) is dictated by the geometry of their loss landscapes, which can be formally analysed by mapping these functions to random fields on manifolds. Instead of VQAs, we can then directly study the corresponding random fields, in our case, the Wishart Hypertoroidal Random Fields (WHRFs). We are mostly interested in the distribution of critical points (especially local minima) of WHRFs. For this purpose, the Kac-Rice formula is presented, reformulated, and simulated. The findings identify a phase transition in the distribution of local minima. Beyond a specific threshold, local minima concentrate near the global minimum in function value, meaning that even local minima are good approximators of the global one. The threshold is governed by the ratio between the problem Hamiltonian degrees of freedom and by the number of independent parameters in the VQA. Since the degrees of freedom parameter scales exponentially, we propose symmetry reduction operations to lower the degrees of freedom. This mathematically reduces the dimension, scaling down the required parameter threshold and enabling to solve bigger problems.


\section{Introduction}

Variational Quantum Algorithms (VQAs), \cite{VQAoverview}, are a promising candidate for demonstrating practical quantum advantage, particularly in quantum chemistry applications such as the ground energy computation for the Fermi-Hubbard model, \cite{Arovas2022,FHSymmetriesBigspatialspin}. While quantum processors can bypass the exponential scaling of the Hilbert space that limits classical computations, near-term hardware remains constrained by noise and decoherence. Although small-scale systems have been successfully solved, \cite{ FermiHub2D, FermiHub3D}, scaling to larger systems exposes a limitation, which is the non-convex topology of the resulting loss landscape. Their complicated nature often makes classical optimisation techniques inefficient.

To scale VQAs effectively, the design of quantum circuit architectures must move beyond inefficient, heuristic trial-and-error approaches. Hamiltonian-agnostic ansatzes navigate the parameter space blindly and often encounter the barren plateau phenomenon or become trapped in suboptimal local traps \cite{Eric_Barren, VQA_learnability}. Overcoming this requires an \textit{a priori} analytical framework that can predict landscape trainability directly from the problem formulation. By formally mapping VQA loss landscapes to Random Fields (RFs) on parameterised manifolds, \cite{Eric_VQA_RF}, this work transforms the quantum hardware problem of trainability into a random matrix theory problem. This mapping identifies a strict phase transition governed by the ratio between the degrees of freedom of the problem Hamiltonian and the number of independent parameters in the VQA. It establishes that an ansatz must cross a specific critical threshold to transition from an underparameterised regime, which is characterised by local minima far from the global minimum in function value, to the overparameterised regime, where local minima concentrate near the global minimum. Effectively, by determining the degrees of freedom of the problem Hamiltonian, we provide an upper estimate for the VQA depth. This is crucial for a reasonable initial choice of the number of blocks in the quantum circuit.

While previous research analytically derived this geometric phase transition, the proofs were established in the asymptotic limit of infinite parameters. To bridge the gap between theoretical results and current hardware reality, this work numerically simulates the discrete Kac-Rice formula to evaluate the random field properties at finite scales. Using the Fermi-Hubbard model as a physical benchmark, our numerical evaluations verify that the phase transition from sub-optimal trapping to global concentration persists for finite systems. We quantify the exponential scaling of the degrees of freedom parameter, demonstrating why standard architectures struggle to reach the trainable state on current quantum hardware. Finally, we demonstrate that by projecting the search space into strict sub-sectors via physical symmetry reductions, the degrees of freedom can be artificially suppressed.

Let us note that the VQA mapping to WHRFs, the Kac-Rice formula derivation, and the Monte Carlo simulations do not inherently alter or improve VQA trainability. Rather, they form a unified statistical framework that models the loss landscape of the VQA as a stochastic process whose associated random matrices follow a Wishart distribution, \cite{Eric_VQA_RF}. These steps provide a theoretical estimate of the required circuit depth, ensuring the model is expressive enough to cross the phase transition without becoming overly parameterised. On the other hand, symmetry reduction is an active intervention on the original problem formulation. By reducing the size of the problem Hamiltonian, the dimension of the resulting hypertoroidal manifold is substantially reduced, thereby lowering the required threshold for successful training across both the VQA and its statistical simulation.

\section{Variational quantum algorithms as Random fields}
\label{sec:VQA_RF}

For the analytical mapping of VQAs to RFs we consider VQAs with the quantum circuits of the following form, \cite{VQAoverview}:

\begin{equation}
\ket{\psi(\theta)} \equiv \displaystyle\prod_{i=1}^{p} U_i(\theta) \ket{\psi_0} \equiv \displaystyle\prod_{i=1}^{p} e^{-i \theta_i Q_i} \ket{\psi_0},
\end{equation}
where $p$ is the number of independent parameterised gates and $\ket{\psi_0}$ is the initial state. We can express the transformations in terms of unitary matrices $U_i(\theta)$ parameterised by $\theta = \{\theta_1, \dots , \theta_p \}$ or, alternatively, using the matrix exponentials, where matrices $Q_i$ are taken directly from the Pauli group $\mathcal{P}_n$.

The parameterised quantum circuit is called an \textit{ansatz} and generates a corresponding loss landscape, which is defined by the cost function, in our case, the energy expectation: 

\begin{equation}
\label{eq:Fvqa_cost_forRF}
    E(\theta) = \bra{\psi(\theta)} H \ket{\psi(\theta)},  
\end{equation}
where $H$ is the problem Hamiltonian and $\ket{\psi(\theta)}$ is the trial state prepared by the ansatz.

In this work, the problem Hamiltonian $H$ represents the Fermi-Hubbard model, \cite{hubbard1963electron, Arovas2022, Somma_2002}. This model simplifies complex many-body systems into a discrete lattice (in our case, a simple 1D lattice). It captures the essential quantum dynamics by simplifying molecular interactions down to two competing mechanisms: kinetic delocalisation and Coulombic repulsion. The total Hamiltonian is therefore governed by two parameters. The kinetic term describes the tunnelling of an electron between adjacent sites with a hopping amplitude $t$. The interaction term represents local electrostatic repulsion, applying an energy penalty $U$ when a single site is doubly occupied by electrons of opposite spin ($\sigma \in \{\uparrow, \downarrow\}$). The full Fermi-Hubbard Hamiltonian in second quantisation is thus defined as:

\begin{equation} \label{eq:Hubbard_Full}
H = -t \sum_{\langle i, j \rangle, \sigma} \left( \hat{a}_{i\sigma}^\dagger \hat{a}_{j\sigma} + \hat{a}_{j\sigma}^\dagger \hat{a}_{i\sigma} \right) + U \sum_{i} \hat{n}_{i\uparrow} \hat{n}_{i\downarrow}
\end{equation}
where $\hat{a}_{i\sigma}^\dagger$ and $\hat{a}_{i\sigma}$ are the standard fermionic creation and annihilation operators, and $\hat{n}_{i\sigma}$ is the number operator.

To evaluate the cost function on a quantum computer, the fermionic Hamiltonian must be mapped to the native qubit hardware. We use the Jordan-Wigner transformation to rewrite the Hamiltonian into qubit operators, \cite{jordan1928paulische, Somma_2002}. The resulting operator is expressed as a sum of local multi-qubit Pauli tensor products:

\begin{equation} 
\label{eq:Hubbard_JW_Mapped}
H = -\frac{t}{2} \sum_{\langle i, j \rangle, \sigma} \left( X_{i\sigma} X_{j\sigma} + Y_{i\sigma} Y_{j\sigma} \right) + \frac{U}{4} \sum_{i} \left( \hat{I} - Z_{i\uparrow} - Z_{i\downarrow} + Z_{i\uparrow} Z_{i\downarrow} \right).
\end{equation}

The goal of the VQA is to update the parameters $\theta$ to minimise the cost function, \eqref{eq:Fvqa_cost_forRF}. Optimising the parameters is very demanding because the loss landscape is usually filled with local minima or experiences the Barren plateau phenomenon, \cite{Eric_Barren, Bittel_2021}. A vast majority of the loss landscape has to be explored, consuming significant computational resources and, in spite of that, yielding suboptimal results, often ending up in a local minimum. 

Rather than attempting full search space exploration, we can analytically evaluate the landscape's trainability by mapping the cost function, \eqref{eq:Fvqa_cost_forRF}, to a Wishart Hypertoroidal Random Field (WHRF). It is crucial to emphasise that this mapping does not directly optimise the VQA parameters, but it can give us a different perspective. By analysing the WHRF's general properties, such as the distribution and clustering of its critical points, we can estimate the required number of independent ansatz parameters.

The formal proof of equivalence between the VQA loss landscape and WHRFs is detailed in \cite{Eric_VQA_RF}. To establish this mapping, the raw energy expectation $E(\theta)$ cannot be used directly, and it must be shifted by the ground state energy $E_0$ (the smallest eigenvalue of $H$). This shifted cost function, $\tilde{E}(\theta) \propto E(\theta) - E_0$, converges in distribution to a WHRF. To define WHRF, we first introduce the Wishart distribution, which is defined over symmetric, positive-semidefinite random matrices, \cite{Stat_Wishart, Stat_Wishart2}.

\begin{definition}
Let $X_1, \dots, X_m$ be independent and identically distributed $d$-dimensional random column vectors drawn from a multivariate normal distribution with zero mean and a $d \times d$ symmetric positive-definite covariance matrix $\Sigma$, such that $X_i \sim \mathcal{N}_d(0, \Sigma)$. The $d \times d$ random matrix $J$ defined as:
\begin{equation}
    J  =\sum_{i=1}^m X_i X_i^T
\end{equation}
is said to follow a Wishart distribution with $m$ degrees of freedom and scale matrix $\Sigma$. 
\end{definition}

To extend this distribution to a random field, we restrict the spatial domain to match the geometry of the quantum ansatz. Because the parameters $\theta = (\theta_1, \dots, \theta_p)$ dictate $2\pi$-periodic physical rotation angles of the circuit gates, the search space is constrained to a $p$-dimensional hypertorus $\mathbb{T}^p \cong (\mathbb{S}^1)^{\times p}$. The embedding of this manifold into a vector space is realised by taking the tensor product of the coordinate vectors for each unit circle:

\begin{equation}
\label{eq:Torus}
    \Phi(\theta) = \bigotimes_{j=1}^p \begin{pmatrix} \cos(\theta_j) \\ \sin(\theta_j) \end{pmatrix}.
\end{equation}

Because $\Phi(\theta)$ is constructed via a $p$-fold tensor product of 2-dimensional vectors, it embeds the hypertorus into a Euclidean space of dimension $2^p$, serving as the bridge connecting the quantum circuit to the statistical field. Combining these concepts, the shifted cost function $\tilde{E}(\theta)$ converges to the WHRF defined as:

\begin{equation}
\label{eq:WHRF}
F_{\text{WHRF}}(\theta) = \Phi(\theta)^T \left( \frac{1}{m}J \right) \Phi(\theta),
\end{equation}

\noindent
where $J \sim \mathcal{W}(m, I_{2^p})$ is a $2^p \times 2^p$ Wishart random matrix normalized by its $m$ degrees of freedom. In the VQA setting, this quadratic form effectively assigns a positive semi-definite random scalar-valued variable to every coordinate on the manifold.

To build an intuition for this mapping, consider how an ansatz operates. Quantum measurement is inherently probabilistic, meaning that for any specific choice of parameters $\theta$, the circuit outputs a probability distribution over the measurement basis states. Thus, we can conceptually treat the entire parameterised ansatz as a sophisticated random variable. Wishart random fields are statistically structured to mimic this exact type of behaviour. In the quantum setting, $\theta$ dictates the physical rotation angles of the circuit gates. Because quantum rotations are periodic, where an angle of $2\pi$ brings the system back to its starting state, the hypertorus emerges as the ideal geometric representation.

\subsection{Degrees of freedom}
\label{sec:DegreesOfFreedom}

To fully define the WHRF mapping for the Fermi-Hubbard model, we must determine the degrees of freedom parameter used to normalise the Wishart random matrix \eqref{eq:WHRF}. While the parameters $t$ and $U$ govern the physical interactions in \eqref{eq:Hubbard_JW_Mapped}, translating this system into a statistical random field requires quantifying its overall computational complexity through its spectral properties. As derived in \cite{Eric_VQA_RF}, the effective degrees of freedom parameter $m$ connects the energy spectrum of the problem Hamiltonian to the WHRF, and is formally defined as:

\begin{equation} 
\label{eq:paramter_m_def}
    m \equiv \frac{||H - \lambda_1 I||_*^2}{||H - \overline{\lambda} I||_F^2},
\end{equation}
where $ H $ is the problem Hamiltonian, $ \lambda_1 $ is the ground state energy (the smallest eigenvalue of $ H $), and $ \overline{\lambda} $ is the mean eigenvalue of $ H $. The numerator consists of the squared nuclear norm ($ ||\cdot||_* $), which sums the singular values of the matrix, while the denominator is the squared Frobenius norm ($ ||\cdot||_F $), representing the sum of the squared matrix elements.

To compute $ m $ for a finite lattice size, we exact-diagonalise the $ 2^n \times 2^n $ qubit Hamiltonian $ H $ (where $ n $ is the number of qubits) to obtain its full spectrum of eigenvalues $ \{\lambda_k\} $. We identify the ground state energy $ \lambda_1 = \min(\lambda_k) $ and calculate the mean eigenvalue $ \overline{\lambda} = \frac{1}{2^n}\sum_k \lambda_k $. Since the shifted Hamiltonian $H - \lambda_1 I$ is positive semi-definite, its nuclear norm is the sum of its eigenvalues. The Frobenius norm of $ H - \overline{\lambda} I $ is the square root of the sum of the squared shifted eigenvalues. Thus, $ m $ is explicitly calculated as:

\begin{equation}
    m = \frac{\left(\sum_{k=1}^{2^n} (\lambda_k - \lambda_1)\right)^2}{\sum_{k=1}^{2^n} (\lambda_k - \overline{\lambda})^2}.
    \label{eq:m_parameter_sums}
\end{equation}

\begin{figure}[h]
    \centering
    \includegraphics[width=0.8\linewidth]{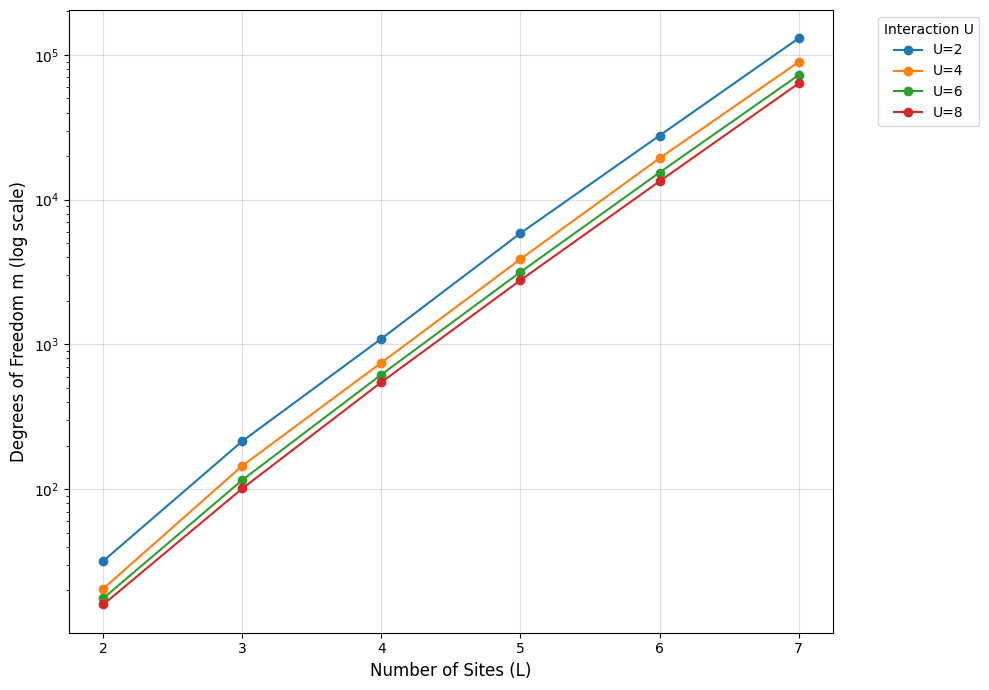}
    \caption{Exponential scaling of the degrees of freedom parameter $m$ with respect to the system size for the Fermi-Hubbard model for varying interaction strengths $U$, with the hopping amplitude fixed at $t=1$ .}
    \label{fig:m_development}
\end{figure}

Physically, the parameter $m$ acts as a signal-to-noise ratio of the Hamiltonian's spectrum, \cite{Eric_VQA_RF}. The numerator represents the total magnitude of the energy deviations from the ground state, while the denominator measures the variance of the eigenvalues. For local Hamiltonians such as the Fermi-Hubbard model, $m$ typically scales exponentially with the number of qubits. As illustrated in \cref{fig:m_development}, this exponential scaling persists across different values of the interaction strength $U$. The hopping amplitude $t$ is fixed to the standard convention of $t=1$, and therefore we can vary only $U$. . Note that the exact calculation of $m$ becomes computationally prohibitive even for modest system sizes due to the $2^N$ dimensionality of the Hamiltonian matrix.

Crucially, this parameter dictates the trainability of the VQAs. By governing the distribution of critical points across the loss landscape, $m$ establishes the minimum number of ansatz parameters required to achieve an expressive, trainable model. Consequently, the exponential scaling of $m$ is a primary factor making the exact optimisation of large-scale VQAs nearly impossible, \cite{QuantumBig}.

\section{Random fields analysis}
\label{sec:RF_analysis}

Having mapped the VQA loss landscape to a WHRF, we analyse its statistical properties to guide the design of efficient ansatz architectures. A central challenge in VQA optimisation is the tendency for algorithms to converge to suboptimal local minima. By examining the distribution of critical points on WHRFs, we quantify the separation between local minima and the global minimum. If local minima concentrate at energy levels distant from the global minimum, the optimisation yields a poor approximation. To formalise the dependence of the landscape topology on the system configuration, we define the overparameterisation factor:

\begin{equation}
\label{eq:OverParameterization}
    \gamma = \frac{p}{2m},
\end{equation}
where $p$ denotes the number of independent trainable parameters in the ansatz (characterising expressivity), and $m$ represents the degrees of freedom of the problem Hamiltonian (characterising problem difficulty). As we demonstrate, reaching a specific threshold in $\gamma$ causes local minima to cluster near the global minimum in function value.

For the analysis of WHRFs, we rely heavily on the Kac-Rice formula, \cite{KacRice_book, KacRice_article}. The following sections provide an intuition and present its formal application to our specific problem, and then detail the numerical sampling techniques used to simulate the WHRF.

\subsection{Kac-Rice formula}
\label{sec:KacRice}

This formula is not derived via any probabilistic tool, but rather arises from a geometric result known as the area formula. This formula can be intuitively derived in this way: For a smooth function $f: \mathbb{R} \to \mathbb{R}$ and a measurable set $T \subseteq \mathbb{R}$, define $N_f(u,T)$, where $u \in \mathbb{R}$, as the number of solutions to $f(t) = u$ with $t \in T$. Then, informally,

\begin{equation}
\label{eq:informalArea}
N_f(u, T) = \int_{f(T)} \delta(v - u) \, dv = \int_T \delta(f(t) - u) \, |f'(t)| \, dt,
\end{equation}
where $\delta$ is the Dirac delta function.

This intuition can be formalised and generalised to higher dimensions. We will leave the proof to \cite{KacRice_book} and only state the formal version of the Area formula.

\begin{definition}[Area formula]
\label{def:Area}
Let $f : U \to \mathbb{R}^d$ be a $\mathcal{C}^1$ function defined on an open subset $U \subset \mathbb{R}^m$, and let $T \subset U$ be a Borel set. Assume that the set of critical values of $f$ has Lebesgue measure zero, and denote $N_f(u, T)$ the number of solutions to the equation $f(t) = u$ with $t \in T$. Then for any Borel-measurable function $g : \mathbb{R}^d \to \mathbb{R}$ that is continuous and bounded,
\begin{equation}
\int_{\mathbb{R}^d} g(u) \, N_f(u, T) \, dy = \int_T |\det f'(t)| \, g(f(t)) \, dt.
\end{equation}
\end{definition}

We can now extend this formula and estimate the number of critical points on a random field. Consider a smooth compact manifold $\mathcal{M}$ of dimension $n$, equipped with a Riemannian metric, and an associated volume measure $\mu_{\mathcal{M}}$. Then, suppose an arbitrary smooth function $f : \mathcal{M} \to \mathbb{R}$. We want to use the Area formula to estimate the moments of the number of critical points of $f$ (we are specifically interested in the first two moments, i.e., expected value and variance). A reasonable hypothesis is to assume that $f$ is almost surely a \emph{Morse} function, i.e., that all its critical points are non-degenerate, \cite{KacRice_book}. Since $\mathcal{M}$ is compact, one can deduce that the number of critical points of a Morse function is finite.

To transition from the general Area Formula to the Kac-Rice formulation, we must specialise the counting measure to find the critical points of the scalar field $f$. A point $w \in \mathcal{M}$ is a critical point if and only if its gradient vanishes. Therefore, we apply the Area Formula by setting our target function to be the gradient vector field: 

\begin{equation}
g(w) = \nabla f(w) = 0.
\end{equation}

Consequently, the Jacobian matrix $g'(w)$, which governs the infinitesimal volume transformation in the Area Formula, evaluates to the derivative of the gradient. This is exactly the Hessian matrix of the original scalar field:

\begin{equation}
g'(w) = \nabla(\nabla f(w)) = \nabla^2 f(w).
\end{equation}

By substituting $g(w) \to \nabla f(w)$ and $g'(w) \to \nabla^2 f(w)$ into the Area Formula integral, the Jacobian determinant $|\det g'(w)|$ is replaced by the Hessian determinant $|\det \nabla^2 f(w)|$.

For any $k \in \mathbb{N}$ and Borel set $B \subseteq \mathbb{R}$, we define $\mathrm{Crit}_{f,k}(B)$ to be the number of critical points $w \in \mathcal{M}$ of $f$ such that $f(x) \in B$ and such that the index of $\mathrm{Hess}\,f(w)$ (that is the number of strictly negative eigenvalues of the Hessian) is at most $k$. The informal area formula of \eqref{eq:informalArea}, applied to $\nabla f$, would read:
\begin{equation}
\mathrm{Crit}_{f,k}(B) = \int_{\mathcal{M}} \delta(\nabla f(w)) |\det \nabla^2 f(w)| \mathbf{1}{\{ f(w) \in B \}} \mathbf{1}\{ \iota(\nabla^2  f(w)) \leq k \}d\mu_{\mathcal{M}}(w).
\end{equation}

Taking the expectation of this equality, one directly obtains the Kac--Rice formula:

\begin{definition}[Kac-Rice formula, informal]
Let $\mathcal{M}$ be a smooth compact Riemannian manifold of dimension $n$, with volume measure $\mu_{\mathcal{M}}$. Let $B \subseteq \mathbb{R}$ be a Borel set. Let $f : \mathcal{M} \to \mathbb{R}$ be an arbitrary function that is almost surely Morse. Denote $p(\nabla f = 0)$ the probability density of $\nabla f(x)$ with respect to the Lebesgue measure on $\mathbb{R}^{n-1}$, taken at $0$. Then:
\begin{equation}
\label{eq:KacRiceInformal}
\begin{split}
\mathbb{E} \, [\mathrm{Crit}_{f,k}(B)] = \\
\int_{\mathcal{M}}  \, \mathbb{E} \Big[  |\det &\nabla^2 f(w)| \, \mathbf{1}\{ f(w) \in B \}  \mathbf{1}\{ \iota(\nabla^2 f(w)) \leq k \} \ \Big| \ \nabla f(w) = 0 \Big] \, p(\nabla f(w) = 0) d {\mu_{\mathcal{M}}(w)}.
\end{split}
\end{equation}
\end{definition}

The rigorous and complete formulation and proof require more assumptions, have to start from a weak form of \eqref{eq:informalArea}, then use continuity arguments to obtain an equality at $u=0$, \cite{KacRice_book}.

The true power of the Kac-Rice formula is that it transforms a problem in differential geometry into a problem of random matrix theory. While this does not automatically make the mathematics trivial, it can unlock different analytical tools for some specific systems. The primary hurdle in applying this formula usually lies in expressing the distribution of the Hessian conditioned on the gradient being zero. Because of this mathematical difficulty, the Kac-Rice formula is usually used for structured random fields, most notably variations of Gaussian fields. Fortunately, WHRFs also possess the structural properties that allow us to apply them.

\subsection{Discrete formulation of Kac-Rice}
\label{sec:KacRiceDiscrete}

The Kac-Rice formula is able to estimate the number of critical points across some interval of function values. The main interest is around the global minimum region. The idea is to compute the number of expected local minima around the global minimum and check whether or not they are clustering around the global minimum. But in order to evaluate the formula, we have to set the range of function values $B$ and then compute the integral over the whole manifold $\mathcal{M}$. This approach is not optimal. Instead of computing the Kac-Rice formula for a range of function values, we show that we can get an estimate for a discrete function value. We then go over a set of function values to get the idea about the whole interval. The integral can be simplified thanks to the invariance of the Wishart distribution. 

In our case, the function on the manifold is \eqref{eq:WHRF}, and we are working with a random field constructed on the hypertorus. The general Kac-Rice expression for computing the number of critical points of index at most $k$ in a set $B$ is written as:

\begin{equation}
\label{eq:KacRice_general}
\mathbb{E}[\mathrm{Crit}_k(B)] 
= \int_{\mathcal{M}} 
\mathbb{E} \Big[ |\det \nabla^2 F| \, \mathbf{1}\{ F \in B \} \, \mathbf{1}\{ \iota (\nabla^2 F) \leq k \} \,\big|\, \nabla F = 0 \Big]
\, p(\nabla F = 0) \, dw.
\end{equation}

We will omit the term $\mathbf{1}\{ \iota (\nabla^2 F) \leq k \}$ for better readability throughout the following transformations. This term can be added at the end without any problem. Also note that we are interested only in $k=0$, because this index corresponds to the local minima we want to study.

Using the law of total probability (also known as the Tower rule), the indicator function $\mathbf{1}\{ F \in B \}$ can be rewritten, and the inner expectation reads as:

\begin{equation}
\mathbb{E}[ |\det \nabla^2 F| \, \mathbf{1}\{ F \in B \} \mid \nabla F = 0 ]
= \int_B \mathbb{E} [ |\det \nabla^2 F|  \mid F = x, \nabla F = 0 ] \, p(F = x \mid \nabla F = 0) \, dx.
\end{equation}

After plugging back into the original equation, \eqref{eq:KacRice_general}, we obtain:
\begin{equation}
\mathbb{E}[\mathrm{Crit}_k(B)] 
= \int_{\mathcal{M}} \int_B 
\mathbb{E} [ |\det \nabla^2 F| \mid F = x, \nabla F = 0 ] \,
p( F = x \mid \nabla F = 0) \, p(\nabla F = 0) \, dx \, dw.
\end{equation}

Evaluating the conditioned probabilities in this form is unfortunately impossible for us. By applying Bayes' theorem to invert the conditioning, substituting the result into \eqref{eq:KacRice_general}, and differentiating with respect to $x$ to isolate the density at a precise function value $F = x$, we obtain the target discrete formulation. Reintroducing the index indicator function $\mathbf{1}\{ \iota (\nabla^2 F) \leq k \}$ yields:

\begin{equation}
\label{eq:KacRiceDiscrFinal}
\mathbb{E} [ \mathrm{Crit}_k(x) ] 
= \int_{\mathcal{M}} 
\mathbb{E} \big[ |\det \nabla^2 F| \, \mathbf{1}\{ \iota (\nabla^2 F) \leq k \} \mid F = x, \nabla F = 0 \big]
\, p(\nabla F = 0 \mid F = x) \, p(F = x) \, dw.
\end{equation}

The main advantage of this formulation is that the terms inside can be directly expressed, resulting in a form that can be effectively simulated, \cite{Eric_VQA_RF}. The distribution of the terms is now provided, starting with the conditioned Hessian.

\begin{theorem}[Distribution of the Hessian conditioned on function value and gradient]
\label{def:condHessian}
Let $F_{\mathrm{WHRF}}(w)$ denote the WHRF defined over the hypertorus $(\mathbb{S}^1)^{\times p}$, and let $x$ denote the value of the random field at a given point. Then, conditioned on $F_{\mathrm{WHRF}}(w) = x$ and $\nabla F_{\mathrm{WHRF}}(w) = 0$, the Hessian $\nabla^2 F_{\mathrm{WHRF}}(w)$ is distributed as, \cite{Eric_VQA_RF}:

\begin{equation}
\label{eq:CondHes}
\left( \nabla^2 F_{\mathrm{WHRF}}(w) \;\middle|\; F_{\mathrm{WHRF}}(w) = x,\, \nabla F_{\mathrm{WHRF}}(w) = 0 \right) \overset{d}{=} \frac{1}{m}(W + \sqrt{2mx}\,G) - 2x I,
\end{equation}
where $W \sim \mathcal{W}_p(2m, I_p)$ is a Wishart matrix with $2m$ degrees of freedom, $G \sim \mathrm{GOE}_p$ is an independent Gaussian Orthogonal Ensemble matrix, and $I$ is the identity matrix.
\end{theorem}

Notice the dimensional difference between the global field matrix $J \in \mathbb{R}^{2^p \times 2^p}$ introduced in \eqref{eq:WHRF} and the matrix $W \in \mathbb{R}^{p \times p}$ in \eqref{eq:CondHes}. While $J$ defines the entire random landscape across the $2^p$-dimensional Euclidean embedding space, the Hessian is evaluated only within the $p$-dimensional tangent space. More details on their relationship can be found in \cite{Eric_VQA_RF}.

The indicator function $\mathbf{1}\{ \iota (\nabla^2 F) \leq 0 \}$ can be introduced back by simply checking if the smallest eigenvalue is greater than zero $\mathbf{1}\{\lambda_1^{\nabla^2F} \geq 0 \}$. 

In a similar way, the distribution of the conditioned gradient can be written, \cite{Eric_VQA_RF}. 

\begin{theorem}[Distribution of the gradient conditioned on function value]
Let $F_{\mathrm{WHRF}}(w)$ denote the WHRF defined over the hypertorus $(\mathbb{S}^1)^{\times p}$, and let $x$ be the value of the random field at a given point. Then, conditioned on $F_{\mathrm{WHRF}}(w) = x$, the gradient $\nabla F_{\mathrm{WHRF}}(w)$ is distributed as

\begin{equation}
(\nabla F_{\mathrm{WHRF}}(w) \,\big|\, F_{\mathrm{WHRF}}(w) = x) 
\;\overset{d}{=} \;
\sqrt{\frac{2x}{m}} \, G,
\end{equation}
where $G$ is GOE distributed and is independent.
\end{theorem}

Consequently, the probability density of the gradient being zero, conditioned on $F_{\mathrm{WHRF}}(w) = x$, is:
\begin{equation}
\label{eq:GradDistr}
p(\nabla F = 0 \mid F = x)
= \frac{1}{\left( 2\pi \frac{2x}{m} \right)^{p/2}}
= \left( \frac{m}{4\pi  x} \right)^{p/2}.
\end{equation}

The probability density function of the function value of $x$, $p(F = x)$, is defined as, \cite{Eric_VQA_RF},

\begin{equation}
\label{eq:GammaDist}
p(F = x) = \frac{m^m}{\Gamma(m)} x^{m-1} e^{-m x}.
\end{equation}

Thanks to the invariance of the Wishart distribution with respect to rotations on the hypertorus, we can integrate out the volume element independently, \cite{Wishart_invariant}:

\begin{equation}
\label{eq:VolumeEl}
\int_{(\mathbb{S}^1)^{\times p}} d w = (2\pi)^p.
\end{equation}

In the context of the Fermi-Hubbard model, the abstract field value $x$ represents the physical energy of the system, $E$. By substituting $x = E$ and combining the expressions for the conditioned Hessian \eqref{eq:CondHes}, the conditioned gradient density \eqref{eq:GradDistr}, the Gamma-distributed field value \eqref{eq:GammaDist}, and the volume element \eqref{eq:VolumeEl}, we obtain the full Kac-Rice expression for the expected number of local minima (critical points of index $k=0$) at a specific energy level:

\begin{equation}
\label{eq:DiscKacRiceFinal}
\begin{aligned}
&\mathbb{E}[\mathrm{Crit}_0(E)] = \\
&\mathbb{E} \Big[ \big| \det \big( \tfrac{1}{m} (W + \sqrt{2mE} G) - 2E I \big) \big| \, \mathbf{1} \big\{ \lambda^{\nabla^2 F}_{1} \ge 0 \big\}\Big]\left( \frac{m}{4\pi  E} \right)^{p/2} \frac{m^m}{\Gamma(m)} E^{m-1} e^{-m E} (2\pi)^p,
\end{aligned}
\end{equation}
where $W \sim \mathcal{W}_p(2m, I)$ is a Wishart matrix, $G \sim \mathrm{GOE}_p$ is a Gaussian Orthogonal Ensemble matrix, $W$ and $G$ are independent, $I$ is the identity matrix, and 
$\lambda^{\nabla^2 F}_{1}$ denotes the smallest eigenvalue of the Hessian.

This expression combines all required components and forms the basis for the practical numerical estimation of the Kac-Rice formula.

\subsection{Simulations}
\label{sec:Simul}

While \eqref{eq:DiscKacRiceFinal} provides the exact analytical expectation for the number of critical points, evaluating the conditioned Hessian determinant requires numerical simulation. As established in \cref{sec:RF_analysis}, the trainability transition is governed by the overparameterisation factor $\gamma$. For a given problem Hamiltonian, the degrees of freedom $m$ remain constant. We therefore hold $m$ fixed and vary the ansatz expressivity $p$ to observe the structural shift in the critical point distribution across uniformly discretised energy levels $E$. 

Equation \eqref{eq:DiscKacRiceFinal} separates into a numerical expectation and an analytical prefactor:

\begin{equation}
\label{eq:KacRiceparts}
\mathbb{E}[\mathrm{Crit}_0(E)] = \underbrace{\mathbb{E} \Big[ \big| \det \big( \tfrac{1}{m} (W + \sqrt{2mE} G) - 2E I \big) \big| \, \mathbf{1} \big\{ \lambda^{\nabla^2 F}_{1} \ge 0 \big\}\Big]}_{\text{Estimation of the conditioned Hessian determinant}}\underbrace{\left( \frac{m}{4\pi  E} \right)^{p/2}\frac{m^m}{\Gamma(m)} E^{m-1} e^{-m E}(2\pi)^p}_{\text{Analytical prefactor}}
\end{equation}

The primary computational bottleneck lies in accurately estimating the expected conditioned Hessian determinant via Monte Carlo sampling. Generating independent samples of this Hessian requires evaluating its exact distribution:

\begin{equation}
\label{eq:HessMatrix}
H = \frac{1}{m} ( W + \sqrt{ 2mE} \, G )- 2  E I.
\end{equation}

\begin{figure}[htbp]
    \centering
    \begin{subfigure}[b]{0.49\textwidth}
        \centering
        \begin{tikzpicture}[
            scale=0.65, every node/.style={scale=0.65},
            node distance=1.6cm,
            process/.style={rectangle, draw, fill=blue!10, text width=5.5cm, text centered, rounded corners, minimum height=3em, font=\small},
            decision/.style={diamond, aspect=2.5, draw, fill=orange!10, text width=3.5cm, text centered, inner sep=0pt, font=\small},
            result/.style={rectangle, draw, fill=green!10, text width=2.5cm, text centered, rounded corners, minimum height=3em, font=\small},
            discard/.style={rectangle, draw, fill=red!10, text width=2.5cm, text centered, rounded corners, minimum height=3em, font=\small},
            arrow/.style={thick,->,>=stealth}
        ]
        
        \node[font=\bfseries\large] (seq_title) at (0, 0) {Sequential Approach};
        \node[process, below of=seq_title, yshift=0.5cm] (seq_gen) {Draw $X_i \in \mathbb{R}^{2m \times p}$, $Y_i \in \mathbb{R}^{p \times p}$};
        \node[process, below of=seq_gen] (seq_mat) {Compute $W_i$, $G_i$ and $H_i$};
        \node[process, below of=seq_mat] (seq_eig) {Compute eigenvalues $\lambda_{i,j}$ of $H_i$};
        \node[decision, below of=seq_eig, yshift=-0.5cm] (seq_cond) {$\min_j(\lambda_{i,j}) > 0$};
        
        \node[result] (seq_true) at ([xshift=-2.2cm, yshift=-2cm]seq_cond.center) {Compute $\det(H_i)$};
        \node[discard] (seq_false) at ([xshift=2.2cm, yshift=-2cm]seq_cond.center) {Set $\det(H_i) = 0$};
        
        \node[process, below of=seq_cond, yshift=-2.5cm] (seq_store) {Store $\det(H_i)$};
        
        \draw[arrow] (seq_gen) -- (seq_mat);
        \draw[arrow] (seq_mat) -- (seq_eig);
        \draw[arrow] (seq_eig) -- (seq_cond);
        
        \draw[arrow] (seq_cond) -| node[anchor=south east, font=\scriptsize] {True} (seq_true);
        \draw[arrow] (seq_cond) -| node[anchor=south west, font=\scriptsize] {False} (seq_false);
        
        \draw[arrow] (seq_true) -- (seq_store);
        \draw[arrow] (seq_false) -- (seq_store);
        
        \draw[thick,->,>=stealth, dashed] (seq_store.west) -- ++(-1.2,0) |- (seq_gen.west) node[pos=0.25, left, font=\scriptsize] {Repeat $N$ times};
        
        \end{tikzpicture}
        \caption{Traditional sequential loop.}
        \label{fig:seq_approach}
    \end{subfigure}
    \hfill
    \begin{subfigure}[b]{0.49\textwidth}
        \centering
        \begin{tikzpicture}[
            scale=0.65, every node/.style={scale=0.65},
            node distance=1.6cm,
            process/.style={rectangle, draw, fill=blue!10, text width=5.5cm, text centered, rounded corners, minimum height=3em, font=\small},
            decision/.style={diamond, aspect=2.5, draw, fill=orange!10, text width=3.5cm, text centered, inner sep=0pt, font=\small},
            result/.style={rectangle, draw, fill=green!10, text width=2.5cm, text centered, rounded corners, minimum height=3em, font=\small},
            discard/.style={rectangle, draw, fill=red!10, text width=2.5cm, text centered, rounded corners, minimum height=3em, font=\small},
            arrow/.style={thick,->,>=stealth}
        ]
        
        \node[font=\bfseries\large] (ten_title) at (0, 0) {Tensor Batching Approach};
        \node[process, below of=ten_title, yshift=0.5cm] (ten_gen) {Draw tensors $X \in \mathbb{R}^{N \times 2m \times p}$, $Y \in \mathbb{R}^{N \times p \times p}$};
        \node[process, below of=ten_gen] (ten_mat) {Batched computation of $\mathbf{W}$, $\mathbf{G}$ and $\mathbf{H}$};
        \node[process, below of=ten_mat] (ten_eig) {Compute batched spectrum $\boldsymbol{\Lambda} \in \mathbb{R}^{N \times p}$};
        \node[decision, below of=ten_eig, yshift=-0.5cm] (ten_cond) {$I_i = \mathbf{1}(\min_j \lambda_{i,j} > 0)$};
        
        \node[result] (ten_true) at ([xshift=-2.2cm, yshift=-2cm]ten_cond.center) {Compute $\det(H_i)$ for valid subset};
        \node[discard] (ten_false) at ([xshift=2.2cm, yshift=-2cm]ten_cond.center) {Set $\det(H_i) = 0$ for indefinite};
        
        \node[process, below of=ten_cond, yshift=-2.5cm] (ten_store) {Store computed determinants};
        
        \draw[arrow] (ten_gen) -- (ten_mat);
        \draw[arrow] (ten_mat) -- (ten_eig);
        \draw[arrow] (ten_eig) -- (ten_cond);
        
        \draw[arrow] (ten_cond) -| node[anchor=south east, font=\scriptsize] {$I_i = 1$} (ten_true);
        \draw[arrow] (ten_cond) -| node[anchor=south west, font=\scriptsize] {$I_i = 0$} (ten_false);
        
        \draw[arrow] (ten_true) -- (ten_store);
        \draw[arrow] (ten_false) -- (ten_store);
        
        \end{tikzpicture}
        \caption{Proposed tensor batching architecture.}
        \label{fig:tensor_approach}
    \end{subfigure}

    \caption{Comparison of sampling workflows for estimating the conditioned Hessian determinant. The traditional sequential loop (a) introduces multiple loops, whereas the proposed tensor approach (b) does everything in one operation.}
    \label{fig:methodology_comparison}
\end{figure}
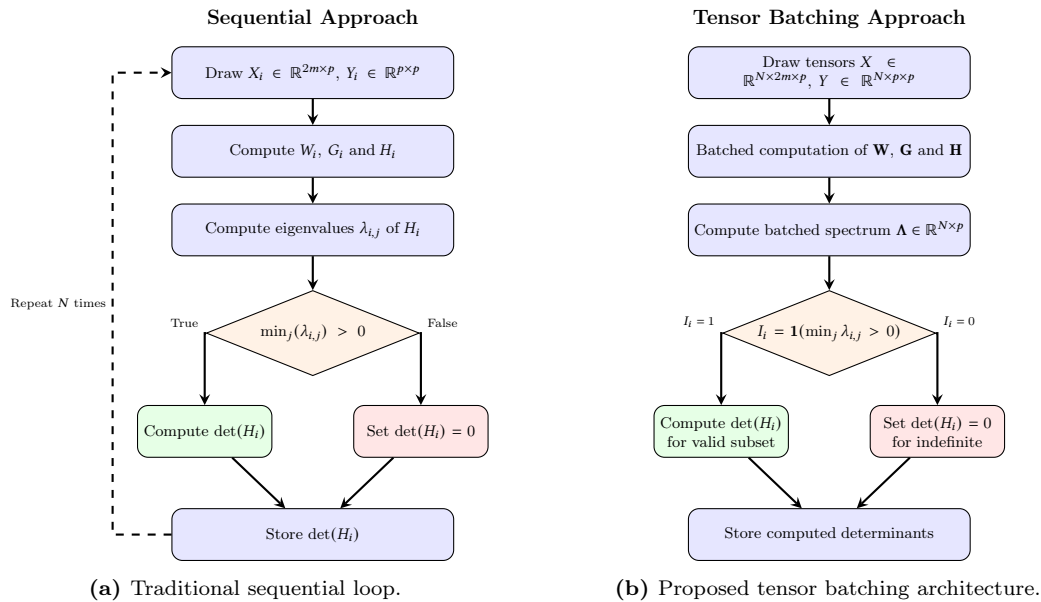

This construction dictates that we first generate independent samples from the Wishart and GOE distributions, which are subsequently scaled and combined to construct the Hessian. Traditionally, Hessians are sampled and processed sequentially, requiring repeated computation of matrix multiplications, eigenvalue decompositions, and determinant computation for each sample, as illustrated in \cref{fig:seq_approach}. The algorithm iterates through each sample $i = 1 \dots N$. In every iteration, it first generates the foundational matrices $X_i \in \mathbb{R}^{2m \times p}$ and $Y_i \in \mathbb{R}^{p \times p}$ by drawing independent standard normal random variables. In the next step, the Wishart matrix $W_i$ and the GOE matrix $G_i$ are constructed:

\begin{equation}
\label{eq:helpingMatrices}
W_i = X_i^\top X_i, \quad G_i = \frac{1}{\sqrt{2}} (Y_i + Y_i^\top).
\end{equation}

Combining these matrices, the conditioned Hessian $H_i$ is constructed, \eqref{eq:HessMatrix}. To enforce the domain restriction, the spectrum  $spec(H_i) = \{\lambda_{i,1}, \dots, \lambda_{i,p}\}$ is computed and the smallest eigenvalue is checked whether or not, it is greater than zero ($\min_j(\lambda_{i,j}) > 0$). If so, the determinant $\det(H_i)$ is calculated and recorded. Otherwise, the determinant is set to zero, $\det(H_i)=0$. 


Although this approach works perfectly for smaller sample sizes, the existence of the loop slows down the process, and we experience troubles when trying to get sample sizes $N>100000$. To accelerate this, we generate \emph{batches} of $N$ samples simultaneously by drawing tensors

\begin{equation}
    X \in \mathbb{R}^{N \times 2m \times p}, \quad Y \in \mathbb{R}^{N \times p \times p},
\end{equation}
where $X$ comprises standard normal variates used to construct Wishart matrices, and $Y$ contains standard normal entries representing GOE matrices. Matrices are then constructed in a similar way as in \eqref{eq:helpingMatrices}.

This enables the construction of all $N$ Hessians in a single operation. The diagonal shift $-2 E I$ is applied efficiently via in-place subtraction, minimising memory overhead.

To identify positive definite Hessians, we determine the spectrum of eigenvalues  $spec(H_i) = \{\lambda_{i,1}, \dots, \lambda_{i,p}\}$ for each matrix in the batch using a standard eigenvalue solver. We then filter for positive definiteness by applying the indicator function:

\begin{equation}
I_i = \mathbf{1}\left(\min_j \lambda_{i,j} > 0\right),
\end{equation}
which isolates the valid subset of matrices. This formulation allows us to efficiently discard indefinite samples prior to the determinant computation. For a visual understanding, see \cref{fig:tensor_approach}.

While this tensor batching significantly accelerates execution by leveraging optimised Basic Linear Algebra Subprograms (BLAS), it inherently trades temporal efficiency for spatial complexity, \cite{dongarra1990set, harris2020array}. The memory footprint of the simulation is bounded by the storage requirements of the foundational tensors $X$ and $Y$, which scale as $\mathcal{O}(Nmp)$ and $\mathcal{O}(Np^2)$, respectively. Consequently, simulating large ensembles to ensure convergence introduces a memory ceiling, practically restricting the maximum allowable parameter count $p$ that can be evaluated simultaneously on standard hardware.

\section{Local minima distribution}
\label{sec:ResDisc}

Previous works studied the distribution of the local minima, specifically in the limit of the problem size. The behavioural shift of the distribution was observed and analytically described. With the Morse theory and free matrix theory, the main result showed that for $p \rightarrow \infty$ the distribution of local minima for energy range $0 \leq E \leq \frac{1}{2}$ has the form of a generalised beta distribution, \cite{Eric_VQA_RF}:

\begin{equation}
\label{eq:GeneralizedDistr}
\mathbb{E}[\mathrm{Crit}_0(E)] \propto e^{-mE} E^{m - \frac{p}{2}} (1 - 2E)^p.
\end{equation}

The behavioural shift is governed by the overparameterisation factor, $\gamma = \frac{p}{2m}$. For $\gamma < 1$, we experience the \textit{underparameterised regime} characterised by local minima concentrating around some finite energy value far from the global minimum. After the threshold is reached and $\gamma > 1$, we enter the \textit{overparameterisation regime} and local minima become concentrated close to the global minimum. 

\begin{figure}[htbp]
    \centering
    \includegraphics[width=1\linewidth]{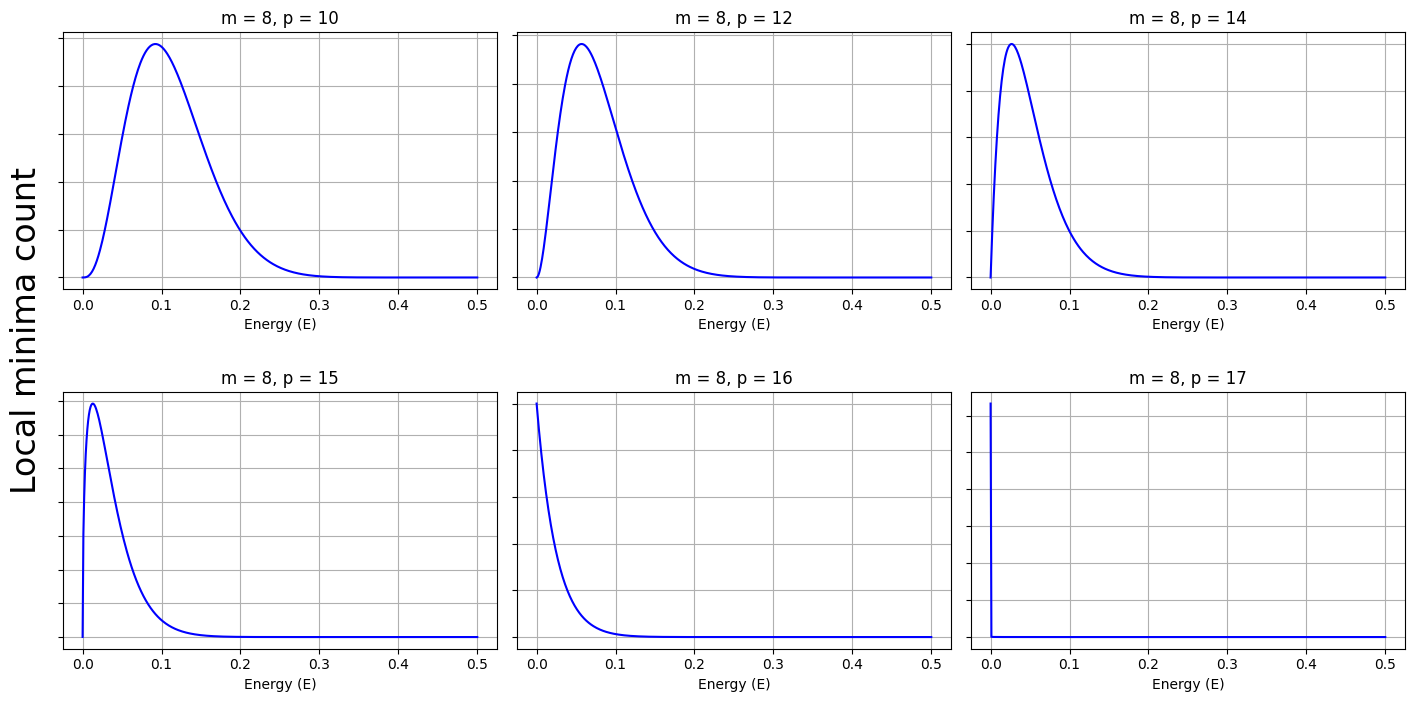}
    \caption{Theoretical transition in the distribution of local minima across the energy domain. Two distinct regimes can be observed. Note that the $y$-axis scales vary across subfigures to best highlight the behavioural change of the local minima.}
    \label{fig:DistrBehav}
\end{figure}

Crucially, local minima start to vanish for $E>0$. To illustrate this, \cref{fig:DistrBehav} plots the distribution defined by \eqref{eq:GeneralizedDistr} for a fixed $m$ and varying $p$. Note that these curves represent unnormalised densities, and the $y$-axis scales vary across subfigures to best highlight the behavioural change of the local minima. We are interested only in this structural behavioural shift, and the exact count of local minima gives us no additional information.

The transition was proved for the case $p \rightarrow \infty$. For smaller model sizes, similar behaviour is expected but not proved analytically. Numerical experiments on VQAs suggest similar changes in behaviour.

\subsection{Discrete Kac-Rice formula simulation}
\label{sec:DiscKacSimul}

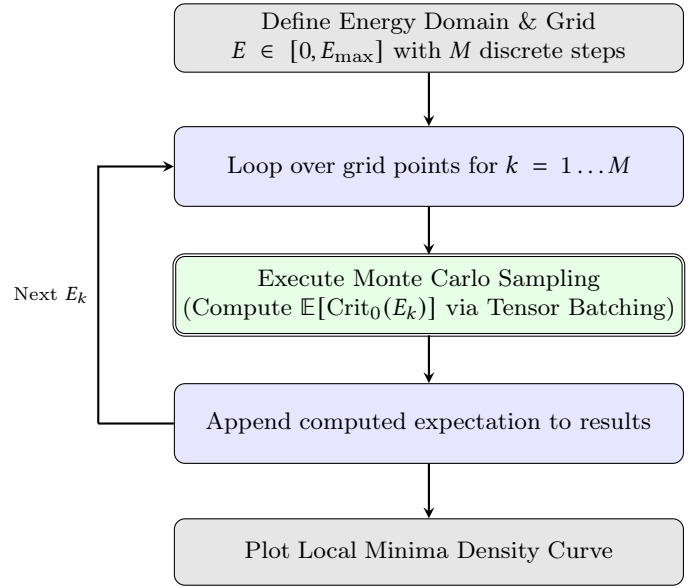
\begin{figure}[htbp]
    \centering
    \begin{tikzpicture}[
        node distance=1.5cm,
        startstop/.style={rectangle, draw, fill=gray!20, text width=6.5cm, text centered, rounded corners, minimum height=2.5em, font=\small},
        process/.style={rectangle, draw, fill=blue!10, text width=6.5cm, text centered, rounded corners, minimum height=3em, font=\small},
        subroutine/.style={rectangle, draw, double, fill=green!10, text width=6.5cm, text centered, rounded corners, minimum height=3em, font=\small},
        arrow/.style={thick,->,>=stealth},
        annotation/.style={rectangle, draw=gray, dashed, fill=gray!5, text width=4cm, text centered, rounded corners, font=\scriptsize}
    ]
    
    \node[startstop] (start) {Define Energy Domain \& Grid\\$E \in [0, E_{\max}]$ with $M$ discrete steps};
    \node[process, below of=start, yshift=-0.2cm] (loop) {Loop over grid points for $k = 1 \dots M$};
    \node[subroutine, below of=loop, yshift=-0.2cm] (core) {Execute Monte Carlo Sampling\\(Compute $\mathbb{E}[\mathrm{Crit}_0(E_k)]$ via Tensor Batching)};
    \node[process, below of=core, yshift=-0.2cm] (agg) {Append computed expectation to results};
    \node[startstop, below of=agg, yshift=-0.2cm] (end) {Plot Local Minima Density Curve};
    
    \draw[arrow] (start) -- (loop);
    \draw[arrow] (loop) -- (core);
    \draw[arrow] (core) -- (agg);
    
    \coordinate (loop_back) at ([xshift=-1cm]agg.west);
    \draw[thick] (agg.west) -- (loop_back);
    \draw[thick,->,>=stealth] (loop_back) |- (loop.west) node[pos=0.25, left, font=\scriptsize] {Next $E_k$};
    
    \draw[arrow] (agg) -- (end);
    
    \end{tikzpicture}
    \caption{Global simulation workflow illustrating the energy domain discretisation, iterative grid evaluation via batched sampling, and final density curve reconstruction.}
    \label{fig:global_workflow}
\end{figure}

To verify that the trainability phase transition persists at finite scales, we evaluate the discrete Kac-Rice formula \eqref{eq:DiscKacRiceFinal} via Monte Carlo simulation. The global simulation workflow, illustrating the discretisation of the energy domain and the nested sampling loop, is detailed in \cref{fig:global_workflow}.

The formula \eqref{eq:DiscKacRiceFinal} partitions into two distinct components, which are a stochastic expectation term and an analytical prefactor. While computing the full distribution requires evaluating both, the fundamental structural shift is entirely captured by the properties of the prefactor. In our approach, we do not study the Kac-Rice formula directly in an analytical way, but by simulating it. The goal is to use the Monte Carlo method to simulate the formula behaviour for finite $p$ model sizes. We would like to verify that this transition also happens for our discrete version of the Kac-Rice formula \eqref{eq:DiscKacRiceFinal}. The formula is partitioned into two distinct components, see \eqref{eq:KacRiceparts}.

While the full distribution requires the evaluation of both, the fundamental phase transition is captured by the properties of the \textit{Analytical prefactor}. The \textit{Estimation of the conditioned Hessian determinant} term acts more like a scaling term. By grouping the constants and separating the terms dependent on energy $E$ in the \textit{Analytical prefactor}, it can be rewritten as:

\begin{equation}
\label{eq:PrefactorSeparated}
P(E) = 
\underbrace{\left[ \left(\frac{m}{4 \pi }\right)^{\frac{p}{2}} \frac{m^m}{\Gamma(m)} (2\pi)^p \right]}_{\mathcal{C}(m, p)} 
\cdot 
\underbrace{E^{m - \frac{p}{2} - 1} e^{-mE}}_{\mathcal{F}(E)}.
\end{equation}

Here, $\mathcal{C}(m,p)$ represents the scaling constants independent of energy, while $\mathcal{F}(E)$ dictates the shape of the distribution. The qualitative behaviour near the global minimum ($E \to 0$) is governed by the exponent $\alpha = m - \frac{p}{2} - 1$, which characterises the analytical competition between the effective degrees of freedom of the problem Hamiltonian $m$ and the number of independent parameters in the ansatz $p$. This exponent acts as a criticality indicator that maps directly to the previously established overparameterisation factor $\gamma$. Specifically, when $\alpha > 0$ (the underparameterised regime), the polynomial term $E^\alpha$ vanishes as $E \to 0$. Conversely, when $\alpha < 0$ (the overparameterised regime), the term $E^\alpha$ diverges at the origin, driving the concentration of local minima exactly at the global minimum. The threshold $\alpha \approx 0$ cleanly marks the boundary where the distribution concentration shifts.

While the \textit{analytical prefactor} governs the phase transition, the \textit{estimation of the conditioned Hessian determinant} dictates the absolute density of local minima and must be evaluated to capture the full distribution. As discussed in \cref{sec:Simul}, accurately estimating this expected value relies on Monte Carlo sampling, which is limited by the positive-definite constraint $\mathbf{1} \{ \lambda^{\nabla^2 F}_{1} \ge 0 \}$. The spectrum of the conditioned Hessian is dependent on the energy level $E$, see \eqref{eq:HessMatrix}.

Across specific regions of the energy spectrum, the probability of drawing a positive-definite matrix from this distribution is suppressed. This transforms the evaluation into a rare event sampling problem. In regions with low acceptance probabilities, independent Monte Carlo sampling yields isolated valid instances, introducing significant statistical variance. This variance manifests as high-frequency noise and artificial density peaks. While this stochastic noise can be mitigated by increasing the sample size $N$, resolving these rare events for bigger problems comes with computational costs.

\subsection{Concrete systems examples}
\label{sec:ConcreteExamples}

\begin{figure}[htbp]
    \centering
    \includegraphics[width=1\linewidth]{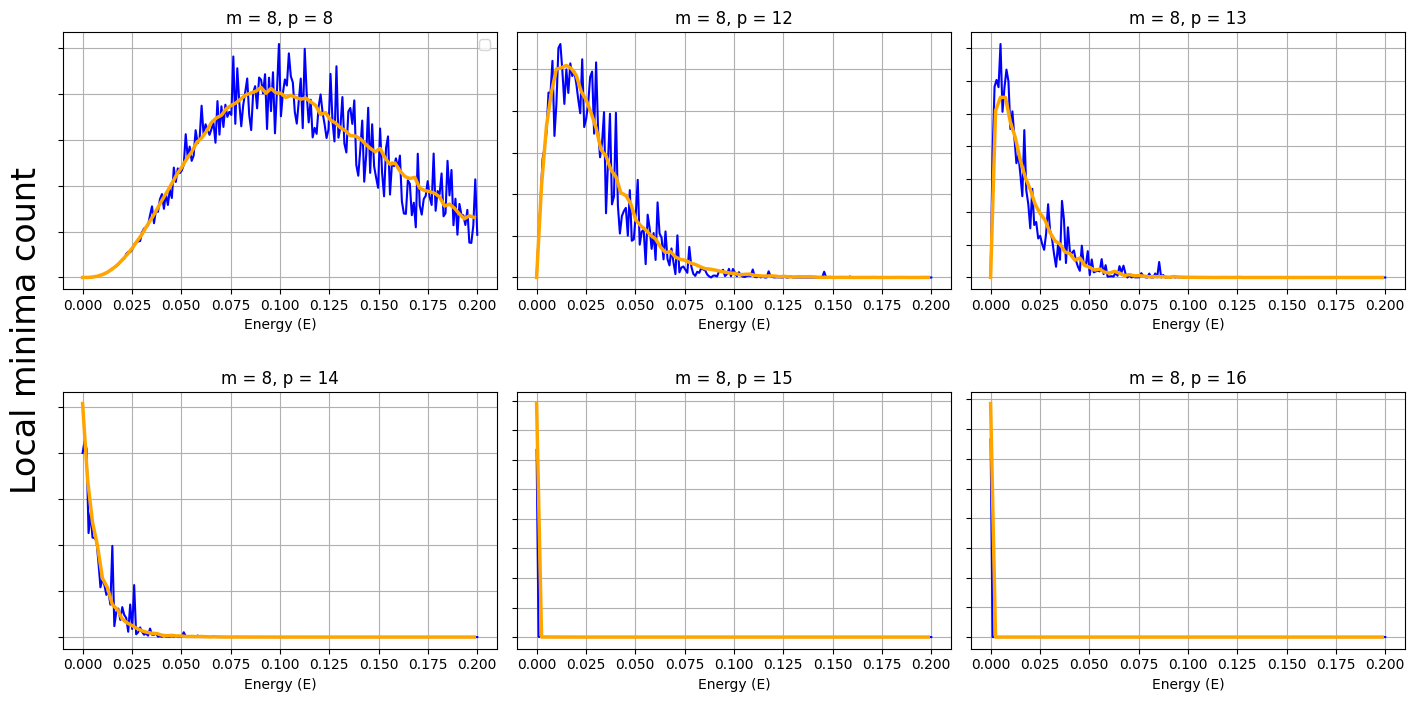}
    \caption{Expected local minima count for the $m=8$ toy model. The plot illustrates both the phase transition across the critical threshold ($p=14$) and the rare-event sampling variance. The blue curves represent a lower sample size ($N=1000$), exhibiting severe stochastic noise, while the orange curves ($N=100,000$) converge to a stable distribution.}
    \label{fig:KacRice_m8_overlayed}
\end{figure}

Having established the theoretical framework, we now evaluate the discrete Kac-Rice formula for concrete systems. To cleanly isolate and examine the underlying mathematical properties of the loss landscape, we first introduce a low-dimensional toy model with the effective degrees of freedom fixed at $m=8$. The parameter $p$ is moved across the range $p \in \{8,12,13,14,15,16\}$ to show the behaviour transition. As anticipated, \cref{fig:KacRice_m8_overlayed} clearly demonstrates that once the critical threshold of $p = 2m - 2 = 14$ is crossed, local minima concentrate around the global minimum. Note that the vertical axis in this and subsequent plots represents the expected local minima count. The exact numerical values are intentionally omitted because the absolute magnitude is unnormalised and provides no additional insight. Our analysis focuses on the qualitative structural shift of the curves. We are especially interested in how the distribution peak migrates as $p$ increases, because this is the true indicator of the phase transition.

To illustrate the rare-event sampling variance, \cref{fig:KacRice_m8_overlayed} contrasts two different sample sizes. The simulation utilising a smaller sample size of $N=1000$ exhibits severe stochastic noise and artificial local fluctuations across the entire energy spectrum. Increasing the sample size for each energy value mitigates these isolated sampling artefacts and gives us a better picture of the distribution change. This example verifies that the shift in the distribution occurs exactly at the predicted threshold. With the same methodology, we can move to physically motivated Hamiltonians.

To test our methodology on a physically motivated problem, we apply it to the 1D Fermi-Hubbard model. Hamiltonian for this system is primarily governed by two parameters: the nearest-neighbour hopping amplitude $t$ and the onsite Coulomb interaction strength $U$. By standard convention, the hopping amplitude is fixed to $t=1.0$ to establish the fundamental energy scale of the system. Consequently, $U$ acts as a dimensionless relative multiplier that determines the ratio of the interaction strength to the kinetic energy. Using this framework, the effective degrees of freedom parameter $m$ was calculated for various system sizes. Results are summarised in \cref{tab:m_values}.

\begin{table}[htbp]
    \centering
    \begin{tabular}{cccc}
        \toprule
        \textbf{Sites} & \textbf{Qubits} & \boldmath{$m \ (U=2)$} & \boldmath{$m \ (U=4)$} \\
        \midrule
        2 & 4  & 32     & 20     \\
        3 & 6  & 215    & 145    \\
        4 & 8  & 1,100  & 749    \\
        5 & 10 & 5,870  & 3,890  \\
        6 & 12 & 27,700 & 19,300 \\
        \bottomrule
    \end{tabular}
    \caption{Numerical evaluation of the parameter $m$ for the Fermi-Hubbard model with hopping parameter $t=1.0$. The parameter $m$ is calculated for varying system sizes (number of sites) and onsite interaction strengths $U$. }
    \label{tab:m_values}
\end{table}

The data in \cref{tab:m_values} demonstrate a rapid, exponential scaling of the complexity as the physical system expands. For example, at an interaction strength of $U=2$, scaling from a trivial 2-site lattice to a modest 6-site lattice causes the degrees of freedom to explode from $m=32$ to $m=27,700$. Recalling that the critical threshold for favourable trainability requires $p > 2m$, a generic ansatz for this 6-site system would demand over 55,000 independent parameters.

Given the parameter scaling for larger systems, we evaluate the discrete Kac-Rice formula on the smallest physically meaningful configuration: a 2-site Fermi-Hubbard lattice simulated on 4 qubits. Based on the numerical evaluations in \cref{tab:m_values}, we focus on the effective degrees of freedom $m=20$ and $m=32$, which correspond to the onsite interaction strengths $U=4$ and $U=2$, respectively. To observe the theoretical phase transition clearly, the parameter $p$ was swept across a range of values specifically chosen to cross the analytical threshold of $p \approx 2m - 2$.

\begin{figure}[htbp]
    \centering
    \includegraphics[width=1\linewidth]{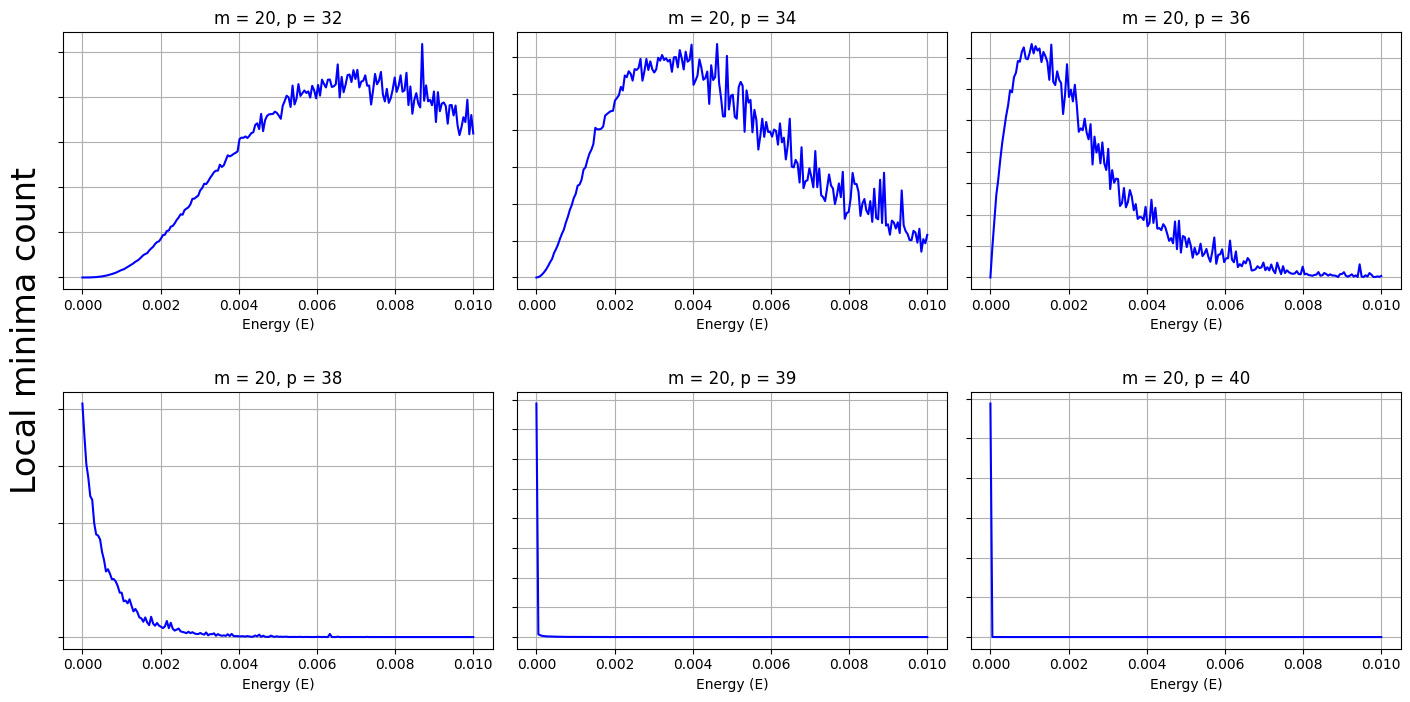}
    \caption{Expected density of local minima for the 2-site Fermi-Hubbard model ($m=20, U=4$). As the number of trainable parameters $p$ crosses the theoretical threshold of $p \approx 38$, the local minima transition from a finite-energy concentration to an exponential concentration at the global minimum.}
    \label{fig:KacRice_m20}
\end{figure}

\begin{figure}[htbp]
    \centering
    \includegraphics[width=1\linewidth]{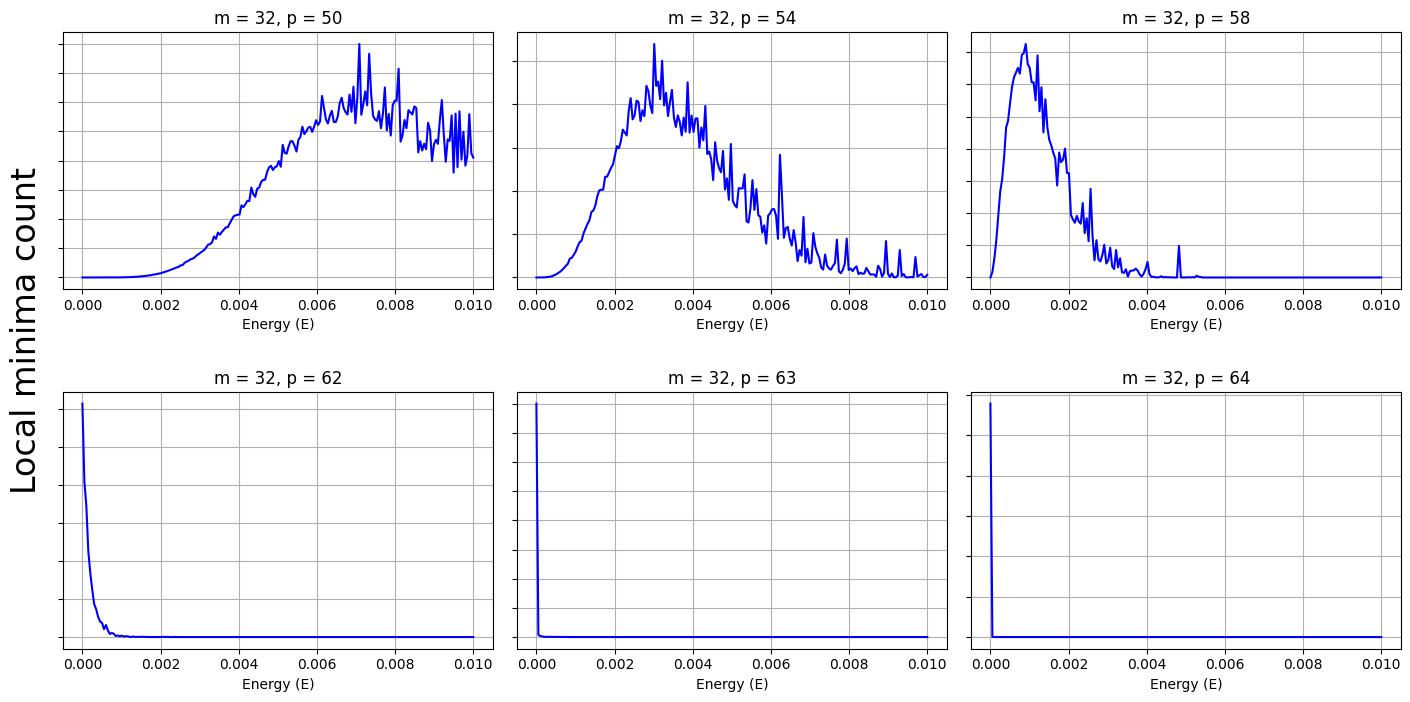}
    \caption{Expected density of local minima for the 2-site Fermi-Hubbard model ($m=32, U=2$), demonstrating the phase transition occurring at the higher threshold of $p \approx 62$.}
    \label{fig:KacRice_m32}
\end{figure}

The resulting distributions of the local minima are presented in \cref{fig:KacRice_m20} and \cref{fig:KacRice_m32}. In these visualisations, we do not show the whole energy range; we instead focus on the critical region adjacent to the global minimum ($E \to 0$). To resolve the rare-event sampling variance, the sampling density of $N=100,000$ is deployed. 

Analysing the progression of the distributions, the behavioural shift is visible. Taking \cref{fig:KacRice_m20} ($m=20$) as the primary example, the theoretical phase boundary is located at $p = 38$. In the underparameterised regime ($p \in \{32, 34,36\}$), the local minima density drops to zero as the energy approaches the global minimum. Instead, the local minima are trapped in a density peak at higher, sub-optimal energy levels. As $p$ increases, this density peak migrates toward the lower energy spectrum. Once the number of trainable parameters reaches the critical threshold ($p=38$), the finite-energy density peak completely vanishes. The local minima exponentially concentrate into a singular peak exactly at the global minimum, ensuring that any found local minimum is an accurate approximator of the global minimum. This same migration is perfectly mirrored in \cref{fig:KacRice_m32} for the $m=32$ system, where the transition occurs as the number of trainable parameters crosses its respective threshold ($p = 62$).

\subsection{Scaling to Medium-Scale Physical Systems}
\label{sec:MediumScale}

The low-dimensional models successfully illustrate the phase transition. But in evaluating bigger models, we face multiple problems. As the degrees of freedom parameter $m$ scales, see \cref{tab:m_values}, our simulations fail. Standard floating-point limits are exceeded, and the memory required to evaluate conditioned Hessian tensors scales exponentially. Therefore, we propose a slightly different strategy to verify the theoretical trends. 

We have implemented a sequential chunking strategy combined with log-space evaluation. In the previous chapter, we were trying to minimise the computational time. We have successfully avoided loops and lowered the time needed. However, drawing a tensor with all samples for an energy level at once becomes impossible because memory limits are exceeded. Therefore, we split the computation into smaller chunks. By generating and evaluating smaller, fixed-size subsets of samples at a time, we bound the memory usage. It is important to note that this technique can help us with the memory wall, but on the other hand, the computational time rises. The next issue we had to deal with was over-floating. This issue was solved by switching to a logarithmic space. This is possible because we are studying the behavioral shift, and therefore, there is no problem in losing the exact local minima count.

\begin{figure}[htbp]
    \centering
    \includegraphics[width=\linewidth]{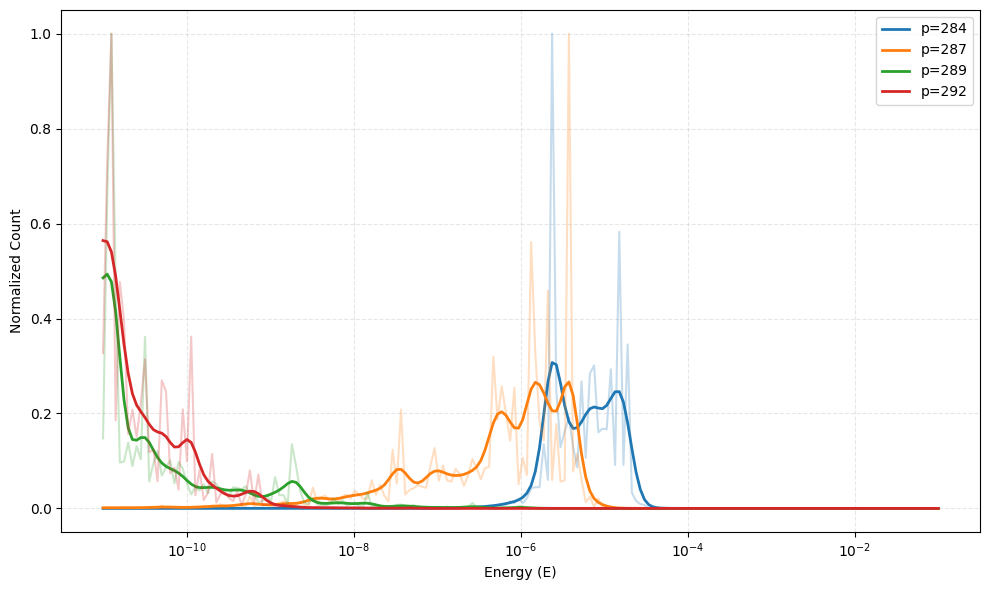}
    \caption{Normalised expected local minima count for the three-site Fermi-Hubbard model ($m=145$, $U=4$). Solid lines represent the distributions smoothed via a 1D Gaussian filter. The raw data is plotted semi-transparently in the background.}
    \label{fig:KacRice_m145}
\end{figure}

Another problem lies in the clear visualisation of our data. The vertical axis continues to represent the expected count of local minima. But since the absolute count of critical points in spaces with hundreds of dimensions grows to large numbers, we max-normalise each curve to a peak value of exactly $1.0$. This normalisation shows the migration of the distribution peak, which is the true indicator of the phase transition. Furthermore, the horizontal energy axis is transformed to a logarithmic scale. This is done because, as the dimensionality of the system increases, the phase transition region compresses close to the global minimum near $E = 0$. A logarithmic scale magnifies this region so the leftward shift of the local minima can be better observed.

Finally, the dimensionality of these physical systems amplifies sampling variance, resulting in heavy fluctuations. In order to observe the macroscopic trend despite these statistical fluctuations, the raw results are plotted semi-transparently in the background. A Gaussian filter is then applied to generate the foreground curves, providing visual clarity for the shifting peaks without hiding the original data. 

\begin{figure}[htbp]
    \centering
    \includegraphics[width=\linewidth]{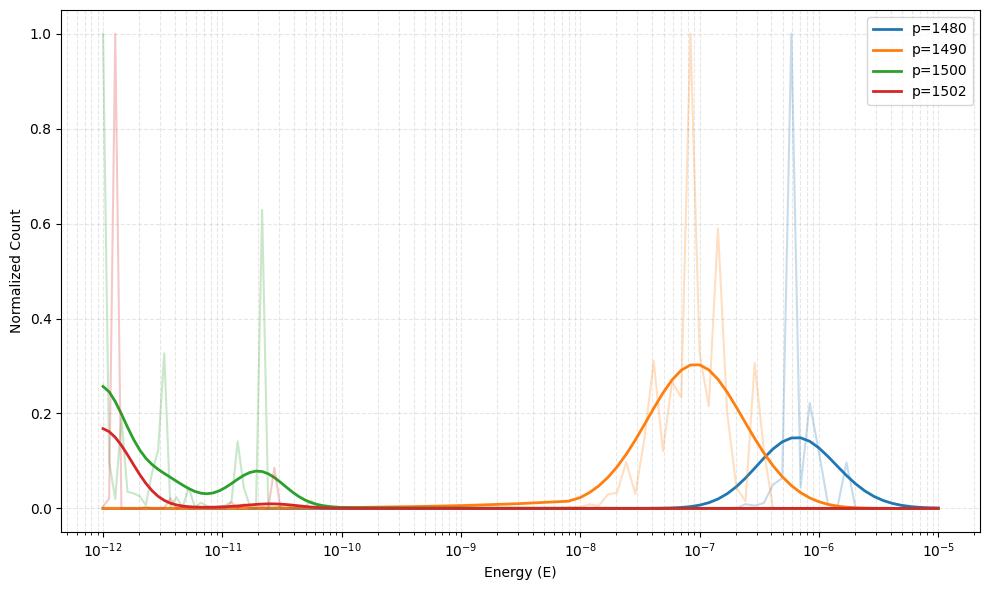}
    \caption{Normalised expected local minima count for the four-site Fermi-Hubbard model ($m=749$, $U=4$). Solid lines represent the distributions smoothed via a 1D Gaussian filter. The raw data is plotted semi-transparently in the background.}
    \label{fig:KacRice_m749}
\end{figure}

With the revised methodology in place, we are able to study bigger systems and verify the behaviour. We evaluate a three-site Fermi-Hubbard model, which has $m=145$ degrees of freedom, see \cref{tab:m_values}. For this specific physical system, the phase transition is theoretically predicted at $p = 2m - 2 = 288$. To capture the evolution across this threshold, we evaluate the discrete Kac-Rice formula for a set of parameters: $p \in \{284, 287, 289, 292\}$. As shown in \cref{fig:KacRice_m145}, the resulting normalised distributions illustrate the theoretical phase transition. In the underparameterized regime ($p < 288$), the local minima are at higher energies. However, when the parameter crosses the threshold, the distribution peaks migrate towards the global minimum at $E \to 0$.

In the next step, we evaluate a four-site Fermi-Hubbard model, which has $m=749$ degrees of freedom, see \cref{tab:m_values}. The phase transition is theoretically predicted at $p = 2m - 2 = 1496$. The transition is again verified, see \cref{fig:KacRice_m749}. At this magnitude, the simulation is quite unstable near the transition phase, and therefore we choose the set of parameters: $p \in \{1480, 1490, 1500, 1502\}$. After the transition, the local minima concentrate at small energies and become hard to capture. This is also the reason for choosing values $p \in \{1500, 1502 \}$. Although we can theoretically verify this behavior for larger $p$, practically we face problems with under-floating. The \cref{fig:KacRice_m749} suffers from heavy fluctuations and is limited in sample size, but still confirms the predicted behaviour. 

While the implementation of a sequential chunking strategy expands the potential scale of the Kac-Rice formula, it does not imply that systems of arbitrarily large dimensions can be simulated given infinite time just by making the chunk size smaller. By evaluating minimal subsets of samples iteratively, our approach makes the memory consumption independent of the total sample size. However, this approach still has to take into consideration the properties of the individual matrices. The absolute ceiling is still bounded by the scaling of a single conditioned Hessian. As the parameter $p$ grows, the memory required to allocate even a single matrix scales quadratically as $\mathcal{O}(p^2)$. 

The analytical proofs guarantee that the theoretical phase transition holds in the asymptotic limit, \cite{Eric_VQA_RF}, and thus scaling these numerical experiments indefinitely is not necessary. And our numerical simulations confirm that the $\gamma = \frac{p}{2m} \approx 1$ trainability rule works not only in the limit, but also for smaller problems. The only minor finite-size discrepancy is that the exact discrete transition occurs slightly earlier, at $p = 2m - 2$. However, because both parameters $p$ and $m$ scale exponentially with the system size, this constant offset of $-2$ vanishes in the limit.

To discuss the computational cost of our numerical experiments, we distinguish between two approaches. For low-dimensional models, we used a full tensor batching approach, which processes big sample sizes (e.g., $N=100,000$) in minutes but suffers from large memory usage that scales as $\mathcal{O}(N p^2)$. On the other hand, for medium-sized systems, the sequential chunking was applied, which saves memory and demands computational time instead. In particular, our hardware configuration was a MacBook Air M3 with 16 GB RAM. The computational times and memory consumption for the cases studied in this paper are presented in \cref{tab:computational_cost}. It is important to note that while the overall runtimes for the medium-scale systems appear comparable to those of the smaller models, this is achieved by reducing the sample size $N$. Keeping the original sample sizes would require days or even weeks of computations.

\begin{table}[htbp]
    \centering
    
    \renewcommand{\arraystretch}{1.2}
    \begin{tabular}{l c c c c c}
        \textbf{Approach} & \textbf{DoF ($m$)} & \textbf{Critical $p$} & \textbf{Samples ($N$)} & \textbf{Peak Memory} & \textbf{Est. Runtime} \\
        \hline
        Tensor Batching & 8 & 14 & 100,000 & $\sim 250$ MB & $\sim 3$ min \\
        & 20 & 38 & 100,000 & $\sim 1.8$ GB & $\sim 1$ h \\
        & 32 & 62 & 100,000 & $\sim 4.5$ GB & $\sim 4.5$ h \\
        \hline
        Sequential Chunking & 145 & 288 & 1,000 & $\sim 300$ MB & $\sim 1.5$ h \\
        & 749 & 1496 & 1,000 & $\sim 2.5$ GB & $\sim 12$ h \\
    \end{tabular}
    
    \caption{Approximate computational demands for evaluating the discrete Kac-Rice formula. Tensor batching and Sequential chunking are compared.}
    \label{tab:computational_cost}
\end{table}

We stress that the choice of simulation parameters, such as sample size, degrees of freedom, number of energy levels and number of $p$ values, significantly affects the runtime and memory consumption. The cases shown in \cref{tab:computational_cost} should illustrate the hardware and runtime demands to the reader.

The numerical analysis of the critical point density and the subsequent identification of the phase transition were conducted using a custom-built Python framework. For the sake of reproducibility, the source code, together with results data, is hosted on GitHub, \cite{Michalek2026VQA}.

\subsection{Implications for VQA Design}
\label{sec:Implications}

The numerical validation of the $p \approx 2m$ trainability threshold presents a challenge for the scalability of VQAs. Because the degrees of freedom $m$ generally scale exponentially with the system size, achieving the overparameterised regime becomes practically impossible.

There are two approaches for the VQAs architecture in the current Noisy Intermediate-Scale Quantum (NISQ) era. Because quantum hardware is still very noisy and suffers from decoherence, the Hardware-Efficient Ansatzes (HEAs) are widely used. They usually take advantage of the gates natively compatible with quantum processors, mitigating the noise and decoherence. These Hamiltonian agnostic architectures work with the assumption that the model is expressive enough and that the classical optimiser will eventually get to the local minimum. However, since HEAs do not discriminate between physically relevant and irrelevant states, they explore the entire parameter space. Consequently, the optimiser evaluates the full, unconstrained Hamiltonian matrix and is subjected to its exponential scaling. As shown in \cref{tab:m_values}, even a minor increase in the physical system size causes the explosion of the space dimension and parameter $m$ subsequently. With current limited quantum hardware capabilities, we inevitably end up deep in the underparameterised regime ($p \ll 2m$) where local minima concentrate at sub-optimal energies.

It is important to note that establishing a direct mapping between the local minima density and actual VQA optimisation performance is difficult. Trainability cannot be resolved simply by increasing $p$ indefinitely. If a quantum circuit becomes too expressive, the loss landscape flattens and suffers, for example, from the barren plateau phenomenon, \cite{Eric_Barren}. Therefore, reducing $m$ via symmetry projections, see \cref{sec:SymmetryReductions}, is important to cross the $\gamma > 1$ threshold without requiring an overly expressive ansatz that may trigger barren plateaus. There are also other elements we have to take into account when discussing the VQA trainability, but in this paper, we focus mainly on the VQA depth.

Another option is to take advantage of the Hamiltonian-Informed Ansatzes (HIAs). Unlike generic HEAs, these informed architectures are constructed using operations that strictly respect the physical properties of the underlying system, such as total particle number or spin parity. If initialised in a physically valid state, an HIA can guarantee that the ansatz will never leave that specific corresponding sub-sector. This approach can help with the trainability problem by "lowering" the parameter $m$. Mathematically, $m$ remains an intrinsic property of the problem Hamiltonian. However, because the HIAs restrict the exploration for specific sub-sectors, it prevents the circuit from accessing states that are physically irrelevant. We can, for example, exclude states that have a different number of particles than the original system, as detailed in \cref{sec:SymmetryReductions}. This mathematically projects the problem down to a much smaller, isolated block of the original Hamiltonian matrix. 

When the degrees of freedom $m$ are calculated strictly within this reduced subspace, their value drops significantly. By shrinking the space VQAs need to explore, the threshold $p \approx 2m$ is lowered, and we can tackle bigger problems. In contrast with the neural networks, where a general architecture can achieve decent results when given enough time and data to be trained, it does not seem to be the same case for VQAs, where we are unable to get good results for general Hamiltonian agnostic architectures.

\section{Symmetry induced reductions of degrees of freedom}
\label{sec:SymmetryReductions}

To overcome the exponential scaling of the full Hilbert space, VQAs must exploit known physical symmetries to restrict the computational search space, \cite{FH_spatial_spin, FH_symmetries_VQA}. In this section, we apply this approach to the 1D Fermi-Hubbard model at half-filling without an external electromagnetic field, quantifying the resulting reduction in the effective degrees of freedom, $m$.

A dynamically conserved physical observable is represented by a Hermitian operator $O$ that commutes with the system's Hamiltonian, $[H, O] = 0$. This commutation ensures that $H$ and $O$ share a common eigenbasis, partitioning the global Hamiltonian matrix into a strict block-diagonal structure, \cite{FH_symm_Book}. Because the Hamiltonian cannot couple states with distinct conserved quantum numbers, transitions between differing symmetry sectors are algebraically forbidden. Consequently, by initialising a VQA in a reference state of a specific symmetry block and employing an ansatz constructed exclusively from commuting generators, the state vector is strictly bound to that physically relevant sub-sector throughout the optimisation process, \cite{FH_symmetries_compound}. 

For the studied unmagnetised 1D Fermi-Hubbard model, we enforce the following symmetries to project the search space:

\begin{itemize}
    \item \textbf{Particle Number Conservation:} The standard Hamiltonian lacks terms that create, annihilate, or flip electrons, independently conserving the total number of spin-up ($N_{\uparrow}$) and spin-down ($N_{\downarrow}$) particles. By enforcing the half-filled, unmagnetised constraint ($N_{\uparrow} = N_{\downarrow} = \lfloor L/2 \rfloor$) the Hilbert space dimension is reduced from $2^{2L}$ to $\binom{L}{\lfloor L/2 \rfloor}^2$, \cite{FH_symmetries_compound, FHSymmetriesBigspatialspin}. Note that for an odd number of lattice sites $L$, a perfectly half-filled state is physically impossible. In this case, we are using the rounded-down integer.
    
    \item \textbf{Spatial Parity ($\mathbb{Z}_2$):} Assuming uniform hopping amplitudes and interactions, the 1D lattice is invariant under spatial inversion. The associated parity operator $P$ satisfies $[H, P] = 0$ and possesses eigenvalues of $\pm 1$. We restrict the search space to the perfectly symmetric ($+1$) eigenspace, effectively halving the remaining dimension, \cite{FH_spatial_spin, FHSymmetriesBigspatialspin}.
    
    \item \textbf{Spin-Flip Parity ($\mathbb{Z}_2$):} In the absence of an external magnetic field, the interaction energy and hopping dynamics are invariant under the global exchange of spin-up and spin-down indices. The spin-flip operator $\Pi_{\text{spin}}$ commutes with the Hamiltonian. For the unmagnetised ground state, we further project the subspace strictly into the symmetric ($+1$) sector of this operator, \cite{FH_SpinFlip}.
\end{itemize}

\subsection{Implications for the degrees of freedom}
\label{sec:ImplicationsForM}

We first note an important theoretical distinction regarding the overparameterisation threshold, $p \approx 2m$. This threshold was derived by mapping the loss landscapes of Hamiltonian Agnostic VQAs to WHRFs. By definition, a Hamiltonian Informed Ansatz restricts the parameterised operations to a specific set of commuting generators, limiting the expressivity of the circuit. Consequently, the random field equivalence may deviate for these restricted models. However, the effective degrees of freedom, $m$, computed on the unconstrained symmetry reduced subspace, serve as an upper bound.

Evaluating \eqref{eq:paramter_m_def} within these restricted subspaces yields the reduced dimensions and corresponding $m$ parameters detailed in \cref{tab:m_reductions}. The table demonstrates the cumulative impact of applying the total particle number, spin parity, and spatial parity constraints. The reduction becomes particularly stark as the system scales. For our largest simulated lattice ($L=7$), the unconstrained Hilbert space spans over $16,000$ dimensions, yielding $m \approx 130,000$. Applying all three symmetries collapses the dimension to $326$, suppressing $m$ to approximately $2,540$.


\begin{table}[h]
    \centering
    \renewcommand{\arraystretch}{1.2}
    \begin{tabular}{@{}ll rrrr@{}}
        \toprule
        \multirow{2}{*}{\textbf{Sites $L$}} & \multirow{2}{*}{\textbf{Metric}} & \multicolumn{4}{c}{\textbf{Symmetry Restriction Level}} \\
        \cmidrule(l){3-6}
        & & \textbf{Full Space} & \textbf{Particle Number} & \textbf{Spin Parity} & \textbf{Spatial Parity} \\
        \midrule
        \multirow{2}{*}{2} 
        & Dim. & 16 & 4 & 3 & 2 \\
        & $m$  & 32.0 & 6.67 & 5.57 & 2.00 \\ 
        \midrule
        
        \multirow{2}{*}{3} 
        & Dim. & 64 & 9 & 6 & 4 \\
        & $m$  & 215 & 22.0 & 14.9 & 8.60 \\ 
        \midrule
        
        \multirow{2}{*}{4} 
        & Dim. & 256 & 36 & 21 & 13 \\
        & $m$  & 1,100 & 160 & 90.2 & 44.7 \\ 
        \midrule
        
        \multirow{2}{*}{5} 
        & Dim. & 1,024 & 100 & 55 & 31 \\
        & $m$  & 5,870 & 533 & 281 & 144 \\ 
        \midrule
        
        \multirow{2}{*}{6} 
        & Dim. & 4,096 & 400 & 210 & 110 \\
        & $m$  & 27,700 & 2,920 & 1,480 & 729 \\ 
        \midrule
        
        \multirow{2}{*}{7} 
        & Dim. & 16,384 & 1,225 & 630 & 326 \\
        & $m$  & 130,000 & 10,100 & 5,050 & 2,540 \\
        \bottomrule
    \end{tabular}
    \caption{Numerical evaluation of the Hilbert space dimension and effective degrees of freedom ($m$) for the 1D Fermi-Hubbard model ($U=2, t=1$). The table demonstrates the cumulative reduction in landscape complexity achieved by sequentially restricting the search space to the relevant symmetries.}
    \label{tab:m_reductions}
\end{table}

\begin{figure}[h]
    \centering
    \includegraphics[width=0.8\linewidth]{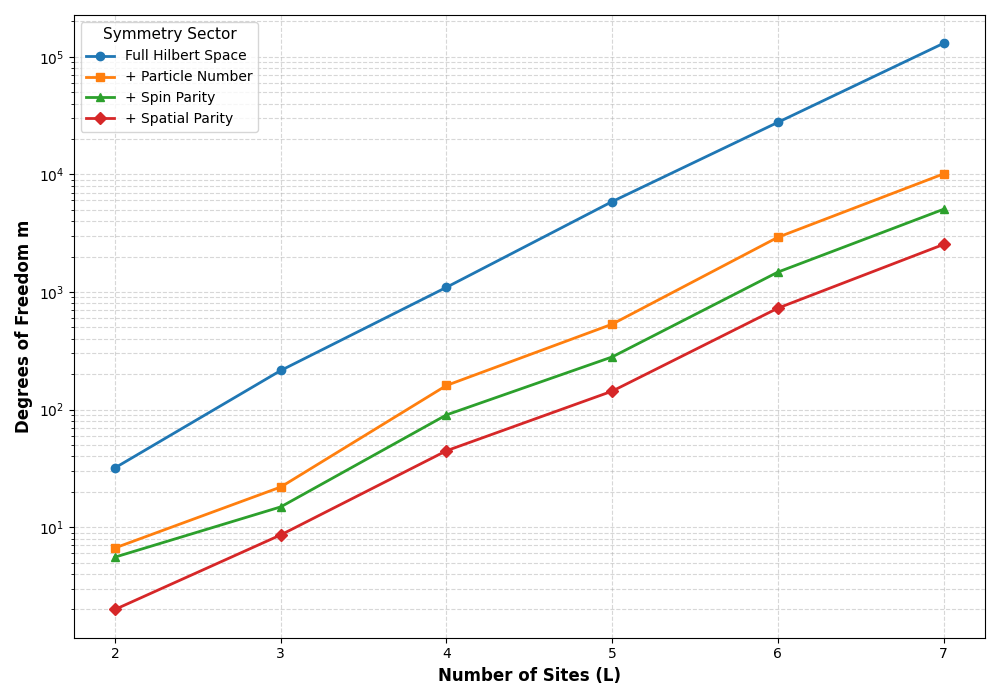}
    \caption{Cumulative reduction of the effective degrees of freedom $m$ through stacked symmetry projections for the 1D Fermi-Hubbard model. Enforcing physical symmetries reduces the absolute landscape complexity compared to the full Hilbert space.}
    \label{fig:m_stacked_symmetries}
\end{figure}

By suppressing $m$, symmetry projections reduce the required number of independent parameters $p$ needed to reach the overparameterisation regime. As illustrated in \cref{fig:m_stacked_symmetries}, the degrees of freedom are significantly reduced compared to the full space. However, the curves confirm that while symmetries can lower the number of dimensions significantly, they do not resolve the fundamental exponential scaling of the Hilbert space.

\section{Conclusion}
\label{sec:Conclustion}

The scalability of Variational Quantum Algorithms (VQAs) is severely limited by non-convex loss landscapes with barren plateaus and suboptimal local minima. To move beyond inefficient trial-and-error hardware design, this work mapped the VQA cost function landscape to Wishart Hypertoroidal Random Fields (WHRFs). By deriving and numerically simulating a discrete formulation of the Kac-Rice formula, we transformed the trainability problem from differential geometry into a numerical simulation of random matrices, allowing us to analyse the distribution of local minima.

The core result is the numerical confirmation of a phase transition in the distribution of local minima governed by the overparameterisation factor $\gamma \approx p/(2m)$. The discrete Kac-Rice simulations demonstrated that when the number of independent parameters ($p$) falls below the critical threshold (for the finite size models $p = 2m - 2$), the system is in an underparameterised regime. In this regime, the probability of a local minimum being close to the global minimum in function value goes to zero, and such a local minimum is a bad approximation. Conversely, once this threshold is crossed into the overparameterised regime, local minima become concentrated near the global one in function, resulting in much better trainability.

Because the degrees of freedom ($m$) scale exponentially with system size, reaching the trainable, overparameterised regime ($p > 2m$) with standard HEAs requires circuit depths that exceed the limitations of current quantum devices. To overcome this scaling barrier, we enforced physical symmetries like conserved particle number, spatial parity, and spin-flip parity. This way, we were able to project the problem Hamiltonian into isolated sub-sectors. This reduces the dimensions of the search space significantly. For instance, applying these symmetry constraints to a 7-site Fermi-Hubbard lattice successfully reduced $m$ from over 130,000 to approximately 2,540, effectively lowering the required parameter threshold to a more realistic number and allowing us to evaluate larger physical systems.

We provide suggestions for VQA design, but certain theoretical boundaries remain open for future exploration. The $p \approx 2m$ critical threshold was mathematically derived by mapping the loss landscapes of Hamiltonian agnostic VQAs to WHRFs. Because HIAs inherently restrict the set of allowable quantum gates to specific symmetry sectors, their exact statistical mapping to random fields may exhibit deviations from the unconstrained Hamiltonian agnostic VQAs. Nevertheless, the restricted degrees of freedom parameter ($m$) remains an upper bound for determining problem difficulty. Future research aims to validate the discrete Kac-Rice threshold predictions against results on actual quantum processors. 

We have already begun testing this approach on the 156-qubit IBM Aachen quantum processor. These preliminary experiments confirmed that hardware constraints must also be taken into account for theoretical trainability estimates. While mathematical symmetry reductions successfully lower the threshold, deploying HIAs introduces multiple compilation problems. Enforcing abstract symmetries often conflicts with the restricted connectivity of physical chips, such as IBM's heavy-hex lattice. Without hardware-specific tailoring, the SWAP-gate routing required to execute these circuits rapidly inflates depth. The transpilation into the hardware native gates also makes the depth interpretation questionable and requires further research.

\section*{Acknowledgement}
This work was supported by the Ministry of Education, Youth and Sports of the Czech Republic under the INTER-EXCELLENCE program (INTER-COST), project no. LUC25028, "Geometric Algebra in Relativistic Quantum Information". This project has been granted based on the COST Action CA23115 "Relativistic Quantum Information". JM and MF also acknowledge financial support from the Czech Academy of Science, in particular the {\it Praemium Academiae} awarded to MF, and the Strategy AV21 program "AI: Artificial Intelligence for Science and Society". We gratefully acknowledge fruitful discussions with MSc. Eric R. Anschuetz, Ph.D.,from the Institute for Quantum Information and Matter \& Walter Burke Institute for Theoretical Physics at Caltech, USA.

\section*{Competing Interests}
All authors declare no financial or non-financial competing interests.

\section*{Data and Code Availability}
The underlying code and datasets for this study are available in the GitHub repository and can be accessed via the following link: \url{https://github.com/HonzaTheGreat/vqa-landscape-geometry}.

\newpage
\bibliographystyle{unsrt}
\bibliography{references}

@inproceedings{
Eric_VQA_RF,
title={Critical Points in Quantum Generative Models},
author={Eric Ricardo Anschuetz},
booktitle={International Conference on Learning Representations},
year={2022},
url={https://openreview.net/forum?id=2f1z55GVQN}
}

@article{Eric_Barren,
   title={Quantum variational algorithms are swamped with traps},
   volume={13},
   ISSN={2041-1723},
   url={http://dx.doi.org/10.1038/s41467-022-35364-5},
   DOI={10.1038/s41467-022-35364-5},
   number={1},
   journal={Nature Communications},
   publisher={Springer Science and Business Media LLC},
   author={Anschuetz, Eric R. and Kiani, Bobak T.},
   year={2022},
   month=dec }

@article{VQA_learnability,
  author    = {Cerezo, M. and Larocca, Martin and García-Martín, Diego and Diaz, N. L. and Braccia, Paolo and Fontana, Enrico and Rudolph, Manuel S. and Bermejo, Pablo and Ijaz, Aroosa and Thanasilp, Supanut and Anschuetz, Eric R. and Holmes, Zoë},
  title     = {Does provable absence of barren plateaus imply classical simulability?},
  journal   = {Nature Communications},
  volume    = {16},
  number    = {1},
  pages     = {7907},
  year      = {2025},
  doi       = {10.1038/s41467-025-63099-6},
  url       = {https://doi.org/10.1038/s41467-025-63099-6}
}

@article{VQAoverview,
  author    = {Cerezo, M. and Arrasmith, Andrew and Babbush, Ryan and Benjamin, Simon C. and Endo, Suguru and Fujii, Keisuke and McClean, Jarrod R. and Mitarai, Kosuke and Yuan, Xiao and Cincio, Lukasz and Coles, Patrick J.},
  title     = {Variational quantum algorithms},
  journal   = {Nature Reviews Physics},
  volume    = {3},
  number    = {9},
  pages     = {625--644},
  year      = {2021},
  doi       = {10.1038/s42254-021-00348-9},
  url       = {https://doi.org/10.1038/s42254-021-00348-9}
}

@article{FermiHub2D,
  author    = {Stanisic, Stasja and Bosse, Jan Lukas and Gambetta, Filippo Maria and Santos, Raul A. and Mruczkiewicz, Wojciech and O'Brien, Thomas E. and Ostby, Eric and Montanaro, Ashley},
  title     = {Observing ground-state properties of the Fermi-Hubbard model using a scalable algorithm on a quantum computer},
  journal   = {Nature Communications},
  volume    = {13},
  number    = {1},
  pages     = {5743},
  year      = {2022},
  doi       = {10.1038/s41467-022-33335-4},
  url       = {https://doi.org/10.1038/s41467-022-33335-4}
}

@article{FermiHub3D,
  title = {Strategies for solving the Fermi-Hubbard model on near-term quantum computers},
  author = {Cade, Chris and Mineh, Lana and Montanaro, Ashley and Stanisic, Stasja},
  journal = {Phys. Rev. B},
  volume = {102},
  issue = {23},
  pages = {235122},
  numpages = {25},
  year = {2020},
  month = {Dec},
  publisher = {American Physical Society},
  doi = {10.1103/PhysRevB.102.235122},
  url = {https://link.aps.org/doi/10.1103/PhysRevB.102.235122}
}

@misc{KacRice_book,
      title={Kac-Rice formula: A contemporary overview of the main results and applications}, 
      author={Corinne Berzin and Alain Latour and José León},
      year={2022},
      eprint={2205.08742},
      archivePrefix={arXiv},
      primaryClass={math.CA},
      url={https://arxiv.org/abs/2205.08742}, 
}

@article{KacRice_article,
  author    = {Nicolaescu, Liviu I.},
  title     = {Counting Zeros of Random Functions},
  journal   = {The American Mathematical Monthly},
  volume    = {130},
  number    = {7},
  pages     = {625--646},
  year      = {2023},
  publisher = {Taylor \& Francis},
  doi       = {10.1080/00029890.2023.2206584}
}

@techreport{Wishart_invariant,
  title       = {Invariant Moments of the Wishart Distribution: A Sage Package and a Website Visualization},
  author      = {Antunes Percíncula, Carlos and Forzani, Liliana and Toledano, Ricardo},
  institution = {Facultad de Ingeniería Química, Universidad Nacional del Litoral},
  year        = {2020},
  type        = {Preprint},
  url         = {https://www.fiq.unl.edu.ar/investigacion/wp-content/uploads/sites/7/2020/12/article.pdf},
  note        = {Submitted to Journal of Statistical Software}
}

@book{Stat_Wishart,
  title={Aspects of Multivariate Statistical Theory},
  author={Muirhead, Robb J.},
  year={2005},
  publisher={John Wiley \& Sons},
  address={Hoboken, NJ},
  note={The definitive reference for the rigorous derivation and properties of the Wishart distribution, including its eigenvalue distributions and non-central forms.}
}

@book{Stat_Wishart2,
  title={An Introduction to Multivariate Statistical Analysis},
  author={Anderson, Theodore W.},
  year={2003},
  edition={3rd},
  publisher={John Wiley \& Sons},
  address={Hoboken, NJ},
  note={Standard foundational text detailing the role of the Wishart distribution in sample covariance matrix theory.}
}

@article{QuantumBig,
  author = {Memon, Q. A. and Al Ahmad, M. and Pecht, M.},
  title = {Quantum Computing: Navigating the Future of Computation, Challenges, and Technological Breakthroughs},
  journal = {Quantum Reports},
  volume = {6},
  number = {4},
  pages = {627--663},
  year = {2024},
  doi = {10.3390/quantum6040039}
}

@book{FH_symm_Book,
  title={The One-Dimensional Hubbard Model},
  author={Essler, Fabian H. L. and Frahm, Holger and G{\"o}hmann, Frank and Kl{\"u}mper, Andreas and Korepin, Vladimir E.},
  year={2005},
  publisher={Cambridge University Press},
  address={Cambridge, UK},
  doi={10.1017/CBO9780511534843}
}

@article{FHSymmetriesBigspatialspin,
  title={The Hubbard model: An introduction and selected rigorous results},
  author={Tasaki, Hal},
  journal={Journal of Physics: Condensed Matter},
  volume={10},
  number={19},
  pages={4353},
  year={1998},
  publisher={IOP Publishing},
  doi={10.1088/0953-8984/10/19/004}
}

@article{FH_SpinFlip,
  title={Two theorems on the Hubbard model},
  author={Lieb, Elliott H},
  journal={Physical Review Letters},
  volume={62},
  number={10},
  pages={1201},
  year={1989},
  publisher={American Physical Society},
  doi={10.1103/PhysRevLett.62.1201}
}

@article{FH_spatial_spin,
  title={Tapering off qubits to simulate fermionic Hamiltonians},
  author={Bravyi, Sergey and Gambetta, Jay M and Mezzacapo, Antonio and Temme, Kristan},
  journal={arXiv preprint arXiv:1701.08213},
  year={2017},
  url={https://arxiv.org/abs/1701.08213}
}

@article{FH_symmetries_VQA,
  title={Exact and approximate symmetry projectors for the electronic structure problem on a quantum computer},
  author={Yen, Tzu-Ching abstract and Lang, Robert A and Izmaylov, Artur F},
  journal={The Journal of Chemical Physics},
  volume={151},
  number={16},
  pages={164111},
  year={2019},
  publisher={AIP Publishing LLC},
  doi={10.1063/1.5110682}
}

@article{FH_symmetries_compound,
  title={Efficient symmetry-preserving state preparation circuits for the variational quantum eigensolver algorithm},
  author={Gard, Bryan T and Zhu, Linghua and Barron, George S and Mayhall, Nicholas J and Economou, Sophia E and Barnes, Edwin},
  journal={npj Quantum Information},
  volume={6},
  number={1},
  pages={1--9},
  year={2020},
  publisher={Nature Publishing Group},
  doi={10.1038/s41534-019-0240-1}
}

@article{hubbard1963electron,
  title={Electron correlations in narrow energy bands},
  author={Hubbard, John},
  journal={Proceedings of the Royal Society of London. Series A. Mathematical and Physical Sciences},
  volume={276},
  number={1365},
  pages={238--257},
  year={1963},
  publisher={The Royal Society}
}

@article{jordan1928paulische,
  title={{\"U}ber das Paulische {\"A}quivalenzverbot},
  author={Jordan, Pascual and Wigner, Eugene},
  journal={Zeitschrift f{\"u}r Physik},
  volume={47},
  number={9-10},
  pages={631--651},
  year={1928},
  publisher={Springer}
}

@misc{Michalek2026VQA,
  author       = {},
  title        = {Source code for VQA landscape analysis via Wishart Hypertoroidal Random Fields},
  year         = {2026},
  publisher    = {GitHub},
  journal      = {GitHub repository},
  howpublished = {\url{https://github.com/HonzaTheGreat/vqa-landscape-geometry}}
}

@article{Bittel_2021,
  title={Training variational quantum algorithms is {NP}-hard},
  author={Bittel, Lennart and Kliesch, Martin},
  journal={Physical Review Letters},
  volume={127},
  number={12},
  pages={120502},
  year={2021},
  publisher={American Physical Society},
  doi={10.1103/PhysRevLett.127.120502}
}

@article{Somma_2002,
  title={Simulating physical phenomena by quantum networks},
  author={Somma, Rolando and Ortiz, Gerardo and Gubernatis, J. E. and Knill, Emanuel and Laflamme, Raymond},
  journal={Physical Review A},
  volume={65},
  number={4},
  pages={042323},
  year={2002},
  publisher={American Physical Society},
  doi={10.1103/PhysRevA.65.042323}
}

@article{Arovas2022,
  title={The Hubbard Model},
  author={Arovas, Daniel P. and Berg, Erez and Kivelson, Steven A. and Raghu, Srinivas},
  journal={Annual Review of Condensed Matter Physics},
  volume={13},
  pages={239--274},
  year={2022},
  publisher={Annual Reviews},
  doi={10.1146/annurev-conmatphys-031620-102024}
}

@article{harris2020array,
  title={Array programming with {NumPy}},
  author={Harris, Charles R and Millman, K Jarrod and van der Walt, St{\'e}fan J and others},
  journal={Nature},
  volume={585},
  number={7825},
  pages={357--362},
  year={2020},
  publisher={Nature Publishing Group}
}

@article{dongarra1990set,
  title={A set of level 3 basic linear algebra subprograms},
  author={Dongarra, Jack J and Du Croz, Jeremy and Duff, Iain S and Hammarling, Sven},
  journal={ACM Transactions on Mathematical Software (TOMS)},
  volume={16},
  number={1},
  pages={1--17},
  year={1990},
  publisher={ACM New York, NY, USA}
}

\end{document}